\documentclass[10pt,journal]{IEEEtran}

\usepackage{amsfonts,amsmath,bm,xspace,caption}
\usepackage{multirow,graphicx, algorithm,algorithmic,rotating}
\usepackage{url,cite}
\usepackage{fullpage}
\usepackage{amsmath,amssymb,graphicx,fullpage,algorithm,algorithmic,bm,amsthm}
\usepackage{color}

\ifCLASSOPTIONcompsoc
\usepackage[caption=false,font=normalsize,labelfont=sf,textfont=sf]{subfig}
\else
\usepackage[caption=false,font=footnotesize]{subfig}
\fi

\usepackage{mathtools}
\DeclarePairedDelimiter{\ceil}{\lceil}{\rceil}
\DeclarePairedDelimiter{\floor}{\lfloor}{\rfloor}

\newcommand{\G}{\mathcal{G}}
\newcommand{\A}{\mathcal{A}}
\newcommand{\N}{\mathcal{N}}
\newcommand{\R}{\mathbb{R}}

\newcommand{\vct}[1]{\bm{#1}}
\newcommand{\mtx}[1]{\bm{#1}}
\newcommand{\vx}{\vct{x}}
\newcommand{\vy}{\vct{y}}

\newcommand{\vv}{\vct{v}}
\newcommand{\va}{\vct{a}}

\newcommand{\vw}{\vct{w}}

\newcommand{\vu}{\vct{u}}

\newcommand{\mA}{\mtx{A}}

\newcommand{\mPhi}{\mtx{\Phi}}

\newcommand{\la}{\langle}
\newcommand{\ra}{\rangle}

\newcommand{\cogent}{CoGEnT\,}
\newcommand{\tol}{\mbox{\rm tol}}

\newtheorem{theorem}{Theorem}[section]

\title{Forward - Backward Greedy Algorithms for Atomic Norm Regularization}

\author{Nikhil Rao \qquad  Parikshit Shah \qquad Stephen Wright  \\
 University of Wisconsin - Madison 
 \thanks{Copyright (c) 2015 IEEE. Personal use of this material is permitted. However, permission to use this material for any other purposes must be obtained from the IEEE by sending a request to pubs-permissions@ieee.org.}}

\date{}

\begin{document}

\maketitle

\begin{abstract}
In many signal processing applications, the aim is to reconstruct a
signal that has a simple representation with respect to a certain
basis or frame. Fundamental elements of the basis known as ``atoms''
allow us to define ``atomic norms'' that can be used to formulate
convex regularizations for the reconstruction problem.  Efficient
algorithms are available to solve these formulations in certain
special cases, but an approach that works well for general atomic
norms, both in terms of speed and reconstruction accuracy, remains to
be found. This paper describes an optimization algorithm called
\cogent that produces solutions with succinct atomic representations
for reconstruction problems, generally formulated with atomic-norm
constraints. \cogent combines a greedy selection scheme based on the
conditional gradient approach with a backward (or ``truncation'') step
that exploits the quadratic nature of the objective to reduce the
basis size. We establish convergence properties and validate the
algorithm via extensive numerical experiments on a suite of signal
processing applications. Our algorithm and analysis also allow for
inexact forward steps and for occasional enhancements of the current
representation to be performed. \cogent can outperform the basic
conditional gradient method, and indeed many methods that are tailored
to specific applications, when the enhancement and truncation steps
are defined appropriately.  We also introduce several novel
applications that are enabled by the atomic-norm framework, including
tensor completion, moment problems in signal processing, and graph
deconvolution.
\end{abstract}


\section{Introduction}
\label{sec:intro}
Minimization of a convex loss function with a constraint on the
``simplicity" of the solution has found widespread applications in
communications, machine learning, image processing, genetics, and
other fields. While exact formulations of the simplicity requirement
are often intractable, it is sometimes possible to devise tractable
formulations via convex relaxation that are (nearly) equivalent.
%
%
%
Since these formulations differ so markedly across applications, a
principled and unified convex heuristic for different notions of
simplicity has been proposed using notions of {\em atoms} and {\em
  atomic norms} \cite{venkat}.  Atoms are fundamental basis elements
of the representation of a signal, chosen so that ``simplicity''
equates to ``representable in terms of a small number of atoms.''  We
list several applications, describing for each application a choice of
atoms that captures the concept of simplicity for those applications.

A sparse signal $x$ may be represented as $x=\sum_{j \in \mathcal{S}}
c_j e_j$, where the $e_j$ are the standard unit vectors and
$\mathcal{S}$ captures the support of $x$. One can view the set
$\left\{ \pm e_j \right\}$ as \emph{atoms} that constitute the signal,
and the convex hull of these atoms is a set of fundamental importance
called the {\em atomic-norm ball}. The operation of
inflation/deflation of the atomic norm ball induces a norm (the
\emph{atomic norm}), which serves as an effective regularizer (see
Sec. \ref{sec:prelim}). The atomic set $\left\{ \pm e_j ,
\; j=1,2,\dotsc,p \right\}$ induces the $\ell_{1}$ norm \cite{CRT},
which is well known to be an effective regularizer for
sparsity. However, this idea can be generalized. For instance, the
atomic norm induced by the convex hull of all unit-rank matrices is
the nuclear norm, often used as a heuristic for rank minimization
\cite{candesmatcomp,rechtmatcomp}. Other novel applications of the
atomic-norm framework include the following.

\begin{itemize}
\item {\bfseries Group-norm-constrained multitask learning} problems
  with group-$\ell_2$ norms \cite{nricip11,bachConsistency,jacob} or
  group-$\ell_{\infty}$ norms \cite{networklinf,negahbanlinf,TurVW05}
  have as atoms unit Euclidean balls and unit $\ell_\infty$-norm
  balls, respectively, restricted to specific groups of variables.
\item {\bfseries Group lasso with overlapping groups} arises from
  applications in genomics, image processing, and machine learning
  \cite{nricip11,jacob}. It is shown in \cite{nraistats} that the sum
  of $\ell_2$ norms of overlapping groups of variables is an atomic
  norm.
\item { \bfseries Moment problems}, which arise in applications such
  as radar, communications, seismology, and sensor arrays, have an
  atomic set which is uncountably infinite \cite{offgrid}. Each atom
  is a trigonometric moment sequence of an atomic measure supported on the unit
  interval \cite{offgrid}.  This methodology can be extended to signal
  classes such as Bessel functions, Gaussians, and wavelets.
\item {\bfseries Group testing on graphs} and network tomography finds
  widespread applications in sensor, computer, social, and biological
  networks \cite{graph_saligrama,jacob}. It is
  typically required to identify a set of faulty edges/nodes from
  measurements that are based on the known structure of the
  graph. Each atom can be defined as a subset of nodes or edges in the
  graph.
\item {\bfseries Hierarchical norms} arise in topic modeling
  \cite{bachhierarchical}, climate and oceanology applications
  \cite{sglclimate}, and fMRI data analysis \cite{soslassonips}. The
  atoms here are hybrids of group-sparse and sparse atoms.
\item {\bfseries OSCAR}-regularized problems use an octagonal penalty
  to simultaneously identify a sparse set of pairwise correlated
  variables \cite{oscar_bondell}. The authors in \cite{mario_oscar}
  show that it can be cast as an atomic-norm-regularized problem. In
  two dimensions, the atoms are the signed canonical basis vectors,
  and vectors with equal magnitude entries.
\item {\bfseries Tensor Completion:} Signals modeled as tensors have
  recently enjoyed renewed interest in machine learning
  \cite{animatensor}. We consider here the case of symmetric,
  orthogonally decomposable, and low-symmetric-rank tensors, in
  which the atoms are unit-rank symmetric tensors.
\item {\bfseries Deconvolution} is the problem of splitting a signal
  $z=x+y$ into two components that are succinct with respect to
  different sets of atoms \cite{mccoydemix}. Typical cases include the
  atomic sets being sparse and low-rank \cite{lowranksparse2}, sparse
  in the canonical and discrete cosine transform (DCT) bases, and
  sparse and group-sparse \cite{dirty}.
\end{itemize}

We present a general method called \cogent (for ``Conditional Gradient
with Enhancement and Truncation'') that can be applied to general
atomic-norm-regularized formulations, including formulations of the
applications discussed above. \cogent minimizes a least-squares loss
function that measures the difference between predictions based on
signal representation and the actual observations, subject to a
``simplicity'' constraint on the signal, imposed via an atomic norm.
Besides its generality, novel aspects of \cogent include (a)
introduction of \emph{enhancement steps} at each iteration to improve
solution fidelity, (b) introduction of efficient \emph{backward steps}
that improve the quality of the reconstruction, (c) introduction of
the notion of \emph{inexactness} in the forward step.



\subsection{Preliminaries and Notation} \label{sec:prelim}


We assume the existence of a known atomic set $\mathcal{A}$ and an
unknown signal $\vx$ in some ``ambient'' space, where $\vx$ is a
superposition of a small number of atoms from $\A$.  (We emphasize
that the set of atoms need not be finite.)
We assume further that the set $\A$ is symmetric about the origin,
that is, $\va \in \A \Rightarrow - \va \in \A$.  The representation of
$\vx$ as a conic combination of atoms $\va \in \mathcal{A}_t$ in a
subset $\A_t \subset \A$ is written as follows:
\begin{equation} \label{eq:co}
\vx= \sum_{\va \in \A_t} \, c_{\va} \va, \;\; \makebox{with $c_{\va} \geq 0$ for
all $\va \in \A_t$}.
\end{equation}
where the $c_{\va}$ are scalar coefficients. We write
\begin{equation}
\label{eq:coA}
\vx \in \mbox{co} (\A_t,\tau),
\end{equation}
for some given $\tau \ge 0$, if it is possible to represent the vector
$\vx$ in the form \eqref{eq:co}, with the additional constraint
\begin{equation} \label{eq:sumco}
\sum_{\va \in \A_t} c_{\va} \le \tau.
\end{equation}
We use $\mA_t $ to denote a linear operator which maps the coefficient
vector $c$ (with cardinality $|\A_t|$) to a vector in the ambient
space, using the vectors in $\A_t$, that is,
\begin{equation} \label{eq:poo1}
\mA_t c := \sum_{\va \in \A_t} c_{\va} \va.
\end{equation}
Since there is a one-to-one relationship between $\A_t$ and the linear
operator $\mA_t$, we use the notation \eqref{eq:poo1} more often, and
sometimes slightly abuse terminology by referring to $\mA_t$ as the
``basis'' at iteration $t$. 
We sometimes refer to the ``columns'' of $\mA_t$, by which we mean the
elements of the corresponding basis $\A_t$.
The \emph{atomic norm} \cite{venkat} is the gauge functional induced
by $\A$:
\begin{equation}
\label{guagedef}
\| \vx \|_\A := \inf \{ t > 0 : \vx \in t (\mbox{conv}({\A}))  \},
\end{equation}
where $\text{conv}\left( \cdot \right)$ denotes the convex hull of a
collection of points.  Equivalently, we have
\begin{equation}
\label{anormdef}
\|\vx\|_{\A} := \inf \left\{ \sum_{\va \in \A} c_a : \vx = \sum_{\va
  \in \A} c_a\va, ~\ c_a \geq 0 
\right\}.
\end{equation} 
Given a representation \eqref{eq:co}, the sum of coefficients in
\eqref{eq:sumco} is an {\em upper bound} on the atomic norm $\| x
\|_{\A}$.
The dual atomic norm is given by
\begin{equation}
\label{dualanorm}
\| \vx \|_\A^* = \sup_{\| \vu \|_\A \leq 1} \la \vu, \vx \ra.
\end{equation}
The dual atomic norm is key to our approach --- the atom selection
step in \cogent (as in the basic CG approach) amounts to choosing the
argument that achieves the supremum in \eqref{dualanorm}, for a
particular choice of $\vx$.

\cogent solves the convex optimization
problem: 
\begin{equation}
\label{eq:opt1}
\min_{\vx} f(\vx):= \frac12 \| \vy - \mPhi \vx
\|^2_2 ~\ \textrm{ s.t. } ~\ \| \vx \|_\A \leq \tau,
\end{equation} 
where $\vy = \mPhi \vx + \vw$ corresponds to observed measurements,
with noise vector $\vw$. The regularizing constraint on the atomic
norm of $\vx$ enforces ``simplicity'' with respect to the chosen
atomic set.  Efficient algorithms are known for this problem when the
atoms are standard unit vectors $\pm \vct{e}_j$ (for which the atomic norm
is the $\ell_1$ norm) \cite{sparsa,troppOMP,tibshirani} and rank-one
matrices (for which the atomic norm is the nuclear norm) \cite{optspace,set}
   \cogent targets the general
formulation \eqref{eq:opt1}, opening up a suite of new applications
with rigorous convergence guarantees and state-of-the-art
empirical performance.

We remark that while \eqref{eq:opt1} is a convex formulation,
efficient algorithms for solving it are not known in full
generality. Indeed, computation of the atomic norm is itself a difficult
operation in some applications. From an optimization perspective,
interior point methods are often impractical, being either difficult
to formulate or too slow for large-scale instances.
By contrast, first-order greedy methods are popular in high
dimensional signal recovery settings because of their computational
efficiency, scalability to large datasets, and interesting global
rate-of-convergence properties. They have found widespread use in
large scale machine learning and signal processing applications
\cite{jaggirevisit,dunn,tewari,harchaoui_gauge,cgnewanal,cosamp,sp}.

\subsection{Past Work: Conditional Gradient Method}

A conditional gradient (CG) algorithm for \eqref{eq:opt1} was
introduced in \cite{tewari}. This greedy approach is often known as
``Frank-Wolfe'' after the authors who proposed it in the 1950s
\cite{frankwolfe}. At each iteration, the CG algorithm finds the atom
that optimizes a first-order approximation to the objective over the
feasible region, and adds this atom to the basis for the
solution. Each iteration of \cogent performs a ``forward step'' of
this type, 
and it is this step that drives the convergence theory, which is
similar to that of standard CG methods \cite{tewari,dunn}, although
some use a different treatment of inexactness in the choice of search
direction. For a detailed review of the CG method, see
\cite{jaggirevisit} and references therein.

Although greedy methods require more iterations than such prox-linear
methods as SpaRSA \cite{sparsa}, FISTA \cite{fista}, and Nesterov's
accelerated gradient method \cite{nesterov_accel}, each iteration is
typically less expensive. For example, in matrix completion
applications, prox-linear methods require computation of an SVD of a
matrix \cite{svt} (or at least a substantial part of it), while CG
requires only the computation of the top singular vector pair. In
other applications, such as structural SVM \cite{lacoste_structSVM},
CG schemes are the only practical way to solve the optimization
formulation. Latent group lasso \cite{jacob} can be extended to
perform regression on very large signals by employing a ``replication"
strategy, but as the amount of group overlap increases, prox-linear
methods quickly become memory intensive. CG offers a scalable
alternative for solving problems of this form.
The procedure to choose each new atom has a linear objective, as
opposed to the quadratic program required to perform projection steps
in prox-linear  methods. The linear subproblem is often easier to
solve, and it 
need
only be solved approximately to retain convergence guarantees
\cite{jaggirevisit,cgnewanal}.

\subsection{Backward (Truncation) Steps} 

In signal processing applications, one is interested not only in
minimizing the loss function, but also in the ``simplicity'' of the
solutions. For example, when the solution corresponds to the wavelet
coefficients of an image, sparsity of the representation is key to its
usefulness as a compact representation.  In this regard, the basic CG
and indeed all greedy schemes suffer from a significant drawback:
Atoms added at some iterations may be superseded by others added at
later iterations, and ultimately may not contribute much to reducing the loss
function.  By the time the loss function has been reduced to an
acceptable level, the basis may contain many such atoms of dubious
value, thus detracting from the quality of the solution.


Backward steps in \cogent allow atoms to be removed from the basis
when they are found to be unhelpful in reducing the objective. We
define this step in a flexible way, the only requirement being that it
does not degrade the objective function too greatly in comparison to
the gain that was obtained at the most recent ``forward'' iteration.
We discuss two possible implementations of this step ---
Algorithms~\ref{alg:bs_multiple} and \ref{alg:bs_matcomp} --- in the
next section.


The ``away steps'' analyzed in \cite{guelat_away} are closely related
to one of our backward/truncation strategies. 
However, the primary consideration in \cite{guelat_away} remains
improvement in the objective value, rather than sparsity: Away steps
are taken only when, to first order, they promise greater decrease in
the objective than the most recently calculated forward step. Because
we seek sparse solutions, we allow backward / truncation steps to be
taken in more general circumstances, even when slight increases in the
objective occur. We note that the ``away steps'' of \cite{guelat_away}
differ from those in the original proposal of Wolfe \cite{Wolfe_away},
which may {\em increase} the size of the basis.

Forward-backward greedy schemes for $\ell_1$ constrained minimization
have been considered previously in \cite{foba_zhang,foba_jain,foba_jalali,LiuFY14a}. These methods build on the Orthogonal Matching Pursuit
(OMP) algorithm \cite{troppOMP}, and cannot be readily extended to the
general setting \eqref{eq:opt1}.

\subsection{Enhancement (Reoptimization) Steps}

\label{sec:enhancement}
The enhancement / reoptimization step in \cogent takes the current
basis and seeks a new set of coefficients in the representation
\eqref{eq:poo1} that reduces the objective while satisfying the norm
constraint. (A ``full correction'' step of this type was described in
\cite{jaggirevisit}.)  The step is implemented as a linear
least-squares objective over the positive orthant of the $\ell_1$ norm
ball. \cogent solves it with a projected gradient method, using a warm
start based on the current set of coefficients.
Since projected gradient is a descent method that maintains
feasibility, it can be stopped after any number of iterations, without
prejudice to the convergence rate of \cogent.  The use of projected
gradient allows ``interpolation'' between the basic CG strategy of
adding the new atom with minimal adjustment of coefficients, and
complete reoptimization over the expanded basis.

\subsection{Outline of  the Paper}

The rest of the paper is organized as follows. We specify \cogent in
the next section, describing different variants of the backward step
that promote parsimonious solutions.
In Section~\ref{sec:analysis}, we state convergence results,
deferring proofs to an appendix.  Section~\ref{sec:exp} describes the
application of \cogent to a number of existing applications, and
compares it to various other methods that have been proposed for these
applications.  In Section~\ref{sec:novelapps}, we apply \cogent for a
variety of {\em new} applications, for which current methods, if they
exist at all, do not scale well to large data sets. In
Section~\ref{sec:decon} we extend our algorithm to deal with
deconvolution problems.

The authors presented a nascent version of \cogent with a few
experimental results in \cite{cogent_icassp}. The algorithm in its
current form was first presented in \cite{cogent_nips}. This paper
explains \cogent in full detail, provides theoretical convergence
guarantees in both exact and approximate settings, and 
extensive empirical results.


\section{Algorithm}
\label{sec:algo}

\cogent is specified in Algorithm~\ref{alg:cogent}. Its three major
elements --- the forward (conditional gradient) step, the backward
(truncation) step, and the enhancement (reoptimization) step --- have
been discussed in Section~\ref{sec:intro}. We note that these three
steps are constructed so that the iterates at each step are feasible
(that is, $\| \vx_t \|_{\A} \leq \tau$).  We make further notes in
this section about alternative implementations of these three steps.




\begin{algorithm}[!ht]
   \caption{CoGEnT: Conditional Gradient with Enhancement and Truncation}
   \label{alg:cogent}
\begin{algorithmic}[1]
   \STATE {\bfseries Input:} Linear oracle for $\A$, bound $\tau$,
    threshold parameter $\eta \in (0,1/2]$;
   \STATE {\bfseries Initialize}, $\va_0 \in
   \A$, $t \leftarrow 0$, $\mA_0 \leftarrow [ \va_0 ]$, $c_0 \leftarrow [\tau ]$, $\vx_0 \leftarrow \mA_0 c_0$;
   \REPEAT 
   \STATE $\va_{t+1} \leftarrow \arg \min_{\va \in \A} \langle \nabla f(\vx_t), \va \rangle;$ \label{greedy}  \COMMENT{FORWARD}
   \STATE $\tilde \mA_{t+1}   \leftarrow  [\mA_t ~\ \va_{t+1}]$;
   \STATE $\gamma_{t+1} \leftarrow \arg \min_{\gamma \in [0,1]} \, f(\vx_t + \gamma (\tau \va_{t+1} - \vx_t))$;  \label{ls} \COMMENT{LINE SEARCH}\\
   \STATE $\tilde{c}_{t+1} \leftarrow [(1 - \gamma_{t+1}) c_t ~\ \gamma_{t+1} \tau \va_{t+1}];$ \label{cgup}
   \STATE \textbf{Optional:} Approximately solve \\
$\tilde{c}_{t+1} \leftarrow \arg \min_{c_{t+1}} f(\tilde \mA_{t+1} c_{t+1}) \; \mbox{s.t. } \, \| c_{t+1} \|_1 \leq \tau, ~\ c_{t+1} \geq 0$ with the output from Step~\ref{cgup} as a warm start; \COMMENT{ENHANCEMENT} \label{enhancement}
   \STATE $\tilde{\vx}_{t+1} = \tilde{\mA}_{t+1} \tilde{c}_{t+1}$;
   \STATE Threshold $F_{t+1} := \eta f(\vx_t) + (1-\eta)
   f(\tilde\vx_{t+1})$;
\STATE $[\mA_{t+1}, c_{t+1}, x_{t+1}]$ \\ $\quad =\mbox{TRUNCATE}(\tilde{A}_{t+1}, \tilde{c}_{t+1}, \tau, F_{t+1})$;  \\
\label{backward} 
\COMMENT{BACKWARD}
   \STATE $t \leftarrow t + 1;$
  \UNTIL{{\bfseries convergence}}
  \STATE {\bfseries Output:} $\vx_t$
\end{algorithmic}
\end{algorithm}


The forward step (Step~\ref{greedy}) is equivalent to solving an
approximation to \eqref{eq:opt1} based on a linearization of $f$
around the current iterate. Specifically, it is easy to show that
$\tau \va_t$ solves the following problem:
\[
\min_{\vx} \, f(\vx_t) + \langle  \nabla f(\vx_t), \vx -\vx_t \rangle
~\ \textrm{ s.t. } ~\ \| \vx \|_{\A} \leq \tau.
\]
(A simple argument reveals that the minimizer of this problem is
attained by $\tau \va$, where $\va$ is an atom.)  We assume the
knowledge of a linear oracle that, given a set of atoms $\A$, returns
the solution of $\arg \min_{a \in \A} \langle \vv , a \rangle$ for a
vector $\vv$. For most applications of interest, the linear oracle can
be calculated efficiently. 

%

The line search of Step~\ref{ls} can be performed exactly, because of
the quadratic objective in \eqref{eq:opt1}. We obtain
\[
\gamma_{t+1} = \min \left\{ \frac{\langle \vy - \mPhi \vx_t , \mPhi
  \vv \rangle}{\| \mPhi \vv \|^2} , 1\right\} , \, \vv := \tau \va_{t+1} - \vx_t.
\]

As mentioned in Section~\ref{sec:enhancement}, Step \ref{enhancement}
can be solved using projected gradient methods. Projection onto the
$\ell_1$ ball can be performed efficiently \cite{duchi_l1}, in
$O(n_{t+1}\log(n_{t+1}))$ operations, where $n_{t+1}$ is the number of
elements in the current basis $\A_{t+1}$.  Each step of projected
gradient yields descent in the objective, so we can terminate the
process before performing full reoptimization over the current basis.
Although the enhancement step is optional, we include it in all the
experiments described in Section~\ref{sec:exp}.

We now discuss two options for performing the backward (truncation)
step (Step~\ref{backward}), whose purpose is to compactify the
representation of $\vx_t$, without degrading the objective more than a
specified amount. The parameter $\eta$ defines a sufficient decrease
criterion that the modified solution needs to satisfy.  A value of
$\eta$ closer to its upper bound will yield more frequent removal of
atoms and hence a sparser solution, at the expense of more modest
progress per iteration.
  
Our first implementation of the truncation step (see
Algorithm~\ref{alg:bs_multiple}) seeks to purge one or more elements
from the expanded basis $\mA_{t+1}$, using a quadratic prediction of
the effect of removal of each atom.  Removal of an atom $\va$ (and
hence the corresponding coefficient $c_{\va}$) from the current
iterate $\tilde\vx_{t+1}$ in Step~\ref{bs:remove} of
Algorithm~\ref{alg:bs_multiple} results in the following change to the
objective:
\begin{align}
\label{eq:quadap}
f& (\tilde \vx_{t+1} - c_{\va} \va) \\
\notag
& = f(\tilde\vx_{t+1}) - c_{\va}
\langle \nabla f(\tilde\vx_{t+1}), \va \rangle + \frac12 c_{\va}^2 \|
\mPhi \va \|_2^2.
\end{align} 
(We have assumed that $c_{\va}$ is the coefficient of $\va$ in the
current representation of $\tilde\vx_{t+1}$.) The scalar quantities
$\| \mPhi \va \|_2^2$ can be computed efficiently and stored as soon
as each atom $\va$ enters the current basis $\mA_{t}$, so the main
cost in evaluating this criterion is in forming the inner product
$\langle \nabla f(\tilde\vx_{t+1}),\va \rangle$.  Having chosen a
candidate atom that optimizes the degradation in $f$, we can simply
remove it from the basis and set its coefficient to zero.
Alternatively, we can reoptimize over the remaining elements
(Step~\ref{eq:betterb} in Algorithm~\ref{alg:bs_multiple}), possibly
using the same projected gradient approach as in
Step~\ref{enhancement} of Algorithm~\ref{alg:cogent}, and test to see
whether the updated value of $f$ still falls below the threshold
$F_{t+1}$.  The projected gradient method in this case will be solving
the following optimization program:
\[
\hat{c}_{t+1} = \arg \min_{c} f(\hat{\mtx{A}}_{t+1}c) ~\ \textbf{s.t.} ~\ c \ge 0, \;
\| c \|_1 \leq \tau.
\]
Atom removal may be repeated in Algorithm~\ref{alg:bs_multiple} as
long as the successively updated objective stays below the threshold
$F_{t+1}$. 

\begin{algorithm}[!ht]
   \caption{: TRUNCATE$(\tilde{A}_{t+1},\tilde{c}_{t+1},\tau,F_{t+1})$}
   \label{alg:bs_multiple}
\begin{algorithmic}[1]
  \STATE {\bf Input:} Current basis $\tilde\mA_{t+1}$, coefficient
  vector $\tilde{c}_{t+1}$, iterate $\tilde{\vx}_{t+1} =
  \tilde{A}_{t+1} \tilde{c}_{t+1}$; bound $\tau$; 
  threshold $F_{t+1}$;
  \STATE continue $\leftarrow 1$;
  \WHILE{continue$=1$}
    \STATE $\hat{\va}_{t+1} \leftarrow \arg \min_{\va \in \tilde\mA_{t+1}} f(\tilde\vx_{t+1} - c_{\va} \va) $ \label{bs:remove}
   \STATE $\hat{\mA}_{t+1} \leftarrow \tilde\mA_{t+1} \backslash \{ \hat{\va}_{t+1}
   \}$;
   \STATE Find $\hat{c}_{t+1} \geq 0$ with $\| \hat{c}_{t+1} \|_1 \le
   \tau$ such that $f(\hat{\mA}_{t+1} \hat{c}_{t+1}) \le f(\tilde\vx_{t+1} -
   (\tilde{c}_{\hat\va_{t+1}})_{t+1} \hat\va_{t+1})$; \label{eq:betterb}  \\
   \IF {$ f(\hat{A}_{t+1} \hat{c}_{t+1}) \le F_{t+1} $}
   \STATE $\tilde\mA_{t+1} \leftarrow \hat{\mA}_{t+1}$; 
   \STATE $\tilde\vx_{t+1} \leftarrow \hat{\mA}_{t+1} \hat{c}_{t+1}$; 
   \STATE $\tilde c_{t+1} \leftarrow \hat{c}_{t+1}$; \\
   \ELSE
   \STATE continue $\leftarrow 0$;
   \ENDIF
\ENDWHILE
   \STATE $\mA_{t+1} \leftarrow \tilde{\mA}_{t+1}$; 
    $\vx_{t+1} \leftarrow \tilde\vx_{t+1}$; 
    $c_{t+1} \leftarrow \tilde{c}_{t+1}$; 

   \STATE {\bf Output:} Possibly reduced basis $\mA_{t+1}$,
   coefficient vector $c_{t+1}\ge 0$, and  iterate
   $\vx_{t+1}$.
\end{algorithmic}
\end{algorithm}


Our second implementation of the truncation step allows for a
wholesale redefinition of the current basis, seeking a new, smaller
basis and a new set of coefficients such that the objective value is
not degraded too much. The approach is specified in
Algorithm~\ref{alg:bs_matcomp}, and is what we use for the matrix and tensor completion experiments later in the paper.  It is motivated by the observation
that atoms added at early iterates contain spurious components, which
may not be canceled out by atoms added at later iterations. This
phenomenon is apparent in matrix completion, where the number of atoms
(rank-one matrices) generated by the procedure above is often
considerably larger than the rank of the target matrix.
For this application, we implement step \ref{svdimplement} of
Algorithm~\ref{alg:bs_matcomp} by forming a singular value
decomposition of the matrix represented by the latest iterate
$\tilde\vx_{t+1}$, and defining a new basis $\hat\mA_{t+1}$ to be a
low-rank matrix that corresponds to the largest singular values. These
singular values would then form the new coefficient vector
$\hat{c}_{t+1}$, and the new iterate $\vx_{t+1}$ would be defined in
terms of just these singular values and singular vectors. The
computational work required for such a step would be comparable with
one iteration of the popular singular value thresholding (SVT)
approach \cite{svt} for matrix completion, which also requires
calculation of the leading singular values and singular vectors. 

\begin{algorithm}[!ht]
   \caption{TRUNCATE$(\tilde{A}_{t+1},\tilde{c}_{t+1},\tau,F_{t+1})$}
   \label{alg:bs_matcomp}
\begin{algorithmic}[1]
  \STATE {\bf Input:} Current basis $\tilde\mA_{t+1}$, coefficient
  vector $\tilde{c}_{t+1}$,  iterate $\tilde{\vx}_{t+1} =
  \tilde \mA_{t+1} \tilde{c}_{t+1}$; bound $\tau$; 
  threshold $F_{t+1}$;
\STATE Find alternative basis $\hat{\mA}_{t+1}$ and coefficients
$\hat{c}_{t+1} \ge 0$ such that $\# columns(\hat{\mA}_{t+1}) < \#
columns(\tilde{\mA}_{t+1})$, $\| \hat{c}_{t+1} \|_1 \le \tau$; \label{svdimplement}
\IF {$f(\hat{\mA}_{t+1} \hat{c}_{t+1}) \le F_{t+1}$}
   \STATE $\mA_{t+1} \leftarrow \hat{\mA}_{t+1}$; 
   $\vx_{t+1} \leftarrow \hat{\mA}_{t+1} \hat{c}_{t+1}$; 
   $c_{t+1} \leftarrow \hat{c}_{t+1}$; \\
   \ELSE
   \STATE $\mA_{t+1} \leftarrow \tilde{\mA}_{t+1}$; 
    $\vx_{t+1} \leftarrow \tilde\vx_{t+1}$; 
    $c_{t+1} \leftarrow \tilde{c}_{t+1}$; \\
   \ENDIF
   \STATE {\bf Output:} Possibly reduced basis $\mA_{t+1}$,
   coefficient vector $c_{t+1}\ge 0$, and  iterate
   $\vx_{t+1}$.
\end{algorithmic}
\end{algorithm}

We conclude this section by discussing practical stopping criteria for
Algorithm~\ref{alg:cogent}.
As we show in Section~\ref{sec:analysis}, \cogent is guaranteed to
converge to an optimum, and the objective is guaranteed to decrease at
each iteration. We therefore use the following termination criteria:
\[
\frac{f(\vx_{t-1}) - f(\vx_t)}{f(\vx_{t-1})} \leq \tol,
\]
where $\tol$ is a small user-defined parameter.

\section{Convergence Results}
\label{sec:analysis}

Convergence properties for \cogent are stated here, with proofs
appearing in the appendix. 
%
Sublinear convergence of \cogent (Theorem \ref{convratenoise}) follows
from a mostly familiar argument. 

\begin{theorem}
\label{convratenoise}
Consider the convex optimization problem \eqref{eq:opt1}, and let
$\vx^*$ be a solution of \eqref{eq:opt1}.  Let $\eta \in
(0,1/2]$. Then the sequence of function values $\{ f(\vx_t) \}$
  generated by \cogent converges to $f^* = f(\vx^*)$ with
\begin{equation} \label{eq:convrate}
f(\vx_T)  - f^* \leq \frac{\bar{C}}{T+1}, \quad \mbox{for all $T \ge 1$},
\end{equation}
where
\begin{align*}
\bar{C}_1 & := \eta D + 2 (1-\eta) LR^2 \tau^2, \\
\bar{C} &:= \frac{2 \bar{C}_1^2}{(1-\eta) (\bar{C}_1 - LR^2 \tau^2)} > 0, \\
L &:= \| \mPhi^T\mPhi \|, \\
D &:= f(\vx_0)-f(\vx^*), \\
R &:= \max_{\va \in \A} \| \va \|.
\end{align*}
\end{theorem}


When the true optimum $\vx^\star$ lies in the interior of the set $\|
\vx \|_\A \leq \tau$, and when $\mPhi$ has full row rank, the
objective function becomes strongly convex and linear convergence
results for the standard CG method apply to \cogent as well
\cite{beck_cgm,guelat_away}. (We omit the formal statement and full
proof of this result, since in most applications of interest, the
solution will lie on the boundary of the atomic-norm ball.)

Similar convergence properties hold when the atom added in the forward
step of Algorithm~\ref{alg:cogent} is computed {\em
  approximately}\footnote{Approximately solving this step can be
  critical in making the approach practical for a wider variety of
  applications, as we see later.}.  In place of the
$\arg\min$ in Step~\ref{greedy} of Algorithm~\ref{alg:cogent}, we have
the following requirement on $\va_{t+1} \in \A$:
\begin{equation}
\label{approxgreedy}
\la \nabla f(\vx_t), \tau \va_{t+1} -\vx_t \ra \le (1-\omega)
\min_{\va \in \A} \la \nabla f(\vx_t), \tau \va -\vx_t \ra
\end{equation}
where $\omega \in (0,1/4)$ is a user-defined parameter.  Unless
$\vx_t$ is a solution, the right-hand side of this expression is
negative, so this condition essentially requires us to find a solution
of the subproblem with relative objective accuracy $\omega$.  
If a tight lower bound for the minimum is available from duality, this
condition can be checked in practice. This criterion requires the
approximate solution to attain a fraction of at least $(1-\omega)$ of
the duality gap, given by $-\min_{\va \in \A} \la \nabla f(\vx_t),
\tau \va -\vx_t \ra$. It is similar in spirit to the inexact Newton
method for nonlinear equations \cite[pp.~277-279]{NW06}, which
requires the approximate solution of the linearized model to achieve
only a fraction of the decrease promised by exact solution of the
model. A similar condition on the relative accuracy of the subproblem
solution was used in \cite[formula~(12)]{lacoste_structSVM} and
\cite[Appendix~A]{Bac13f}.

For the relaxed definition \eqref{approxgreedy} of $\va_{t+1}$, we
obtain the following result.
\begin{theorem}
\label{convrateapprox}
Assume that the conditions of Theorem~\ref{convratenoise} hold, but
that the atom $\va_{t+1}$ selected in Step \ref{greedy} in Algorithm
\ref{alg:cogent} satisfies the condition \eqref{approxgreedy}. Assume
further than $\eta \in (0,1/3)$ and $\omega \in (0,1/4)$. Then we have
\begin{equation} \label{eq:convrate.approx}
f(\vx_T)  - f^* \leq \frac{\tilde{C}}{T+1} \quad \mbox{for all $T \ge 1$},
\end{equation}
where
\begin{align*}
\tilde{C}_1 & := (\eta + \omega(1-\eta))D + 2 (1-\eta) LR^2 \tau^2, \\
\tilde{C}   & := \frac{2 \tilde{C}_1^2}{(1-\eta)[(1-\omega) \tilde{C}_1 - LR^2 \tau^2]},
\end{align*} 
with $L, R, \tau, D$ defined as in Theorem~\ref{convratenoise}
\end{theorem}


\section{Experiments: Standard Applications in Sparse Recovery} 
\label{sec:exp}

\cogent can be used to solve a variety of problems in signal
processing and machine learning, as we show in the remainder of the
paper. In all our experiments (including those in Section
\ref{sec:novelapps}), unless specified otherwise, we choose a random
atom for initialization, and set the parameter $\eta$ to $0.5$.  All
times reported correspond to the time taken for the algorithm to begin
its first iteration and ``finish,'' exhausting the specified maximum
number of iterations or converging to the solution based on the
tolerance value provided. Wherever available, we set $\tau$ to be the
atomic norm of the true signal of interest, and for competing
algorithms, supply the true value of the parameter required by the
method (sparsity, rank, etc.). Finally, in all tables and figures, we
use ``CG'' to refer to the variant of conditional gradient in which
line search is used to find the optimal step size at each iteration.

\subsection{Sparse Signal Recovery}

We first start with the well known case of sparse signal recovery.
We tested our method on the following formulation:
\begin{equation} \label{eq:cs}
\hat{\vx} = \arg \min_{\vx \in \R^p} \| \vy - \mPhi \vx \|_2^2 ~\ \text{ s.t. } \|\vx\|_1 \leq \tau.
\end{equation}
The atoms in this case are the signed canonical basis vectors, and the
atom selection step (Step \ref{greedy} in Algorithm~\ref{alg:cogent})
reduces to the following:
\begin{align*}
\hat{i} & = \arg \max_i | [\nabla f(\vx_t)]_i |, \\
\va_{t+1} & = -\mbox{sign}\left( [\nabla f(\vx_t)]_{\hat{i}} \right)   e_{\hat{i}}.
\end{align*}
The above operation amounts to performing a sort, which can be done in $O(p \log(p))$ time. We consider a sparse signal $\vx$ of length $p = 20000$, with $5\%$ of
randomly set to nonzero values. \footnote{Here and subsequently, the
  phrase ``we randomly set coefficients to be nonzero'' means that we
  select coefficients uniformly and assign them values from the normal
  distribution $\N(0,1)$.}. Setting $n=5000$, we construct the $n
\times p$ matrix $\mPhi$ to have i.i.d. Gaussian entries, and corrupt
the measurements with Gaussian noise (AWGN) of standard deviation
$\sigma = 0.01$.  In the formulation \eqref{eq:cs}, we set $\tau = \|
\vx^\star \|_1$, where $\vx^\star$ is the chosen optimal signal. 

\begin{figure}
\centering 
\includegraphics[width=.24\textwidth]{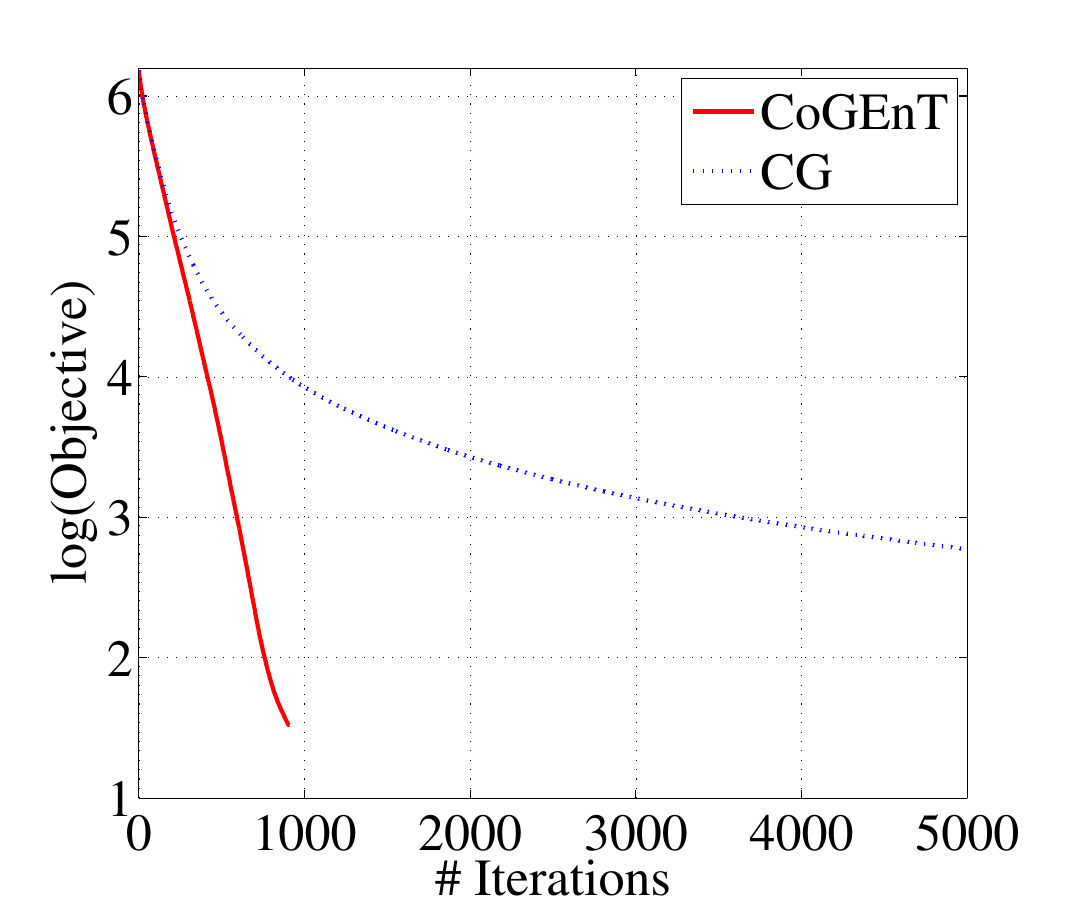}
\includegraphics[width=.24\textwidth]{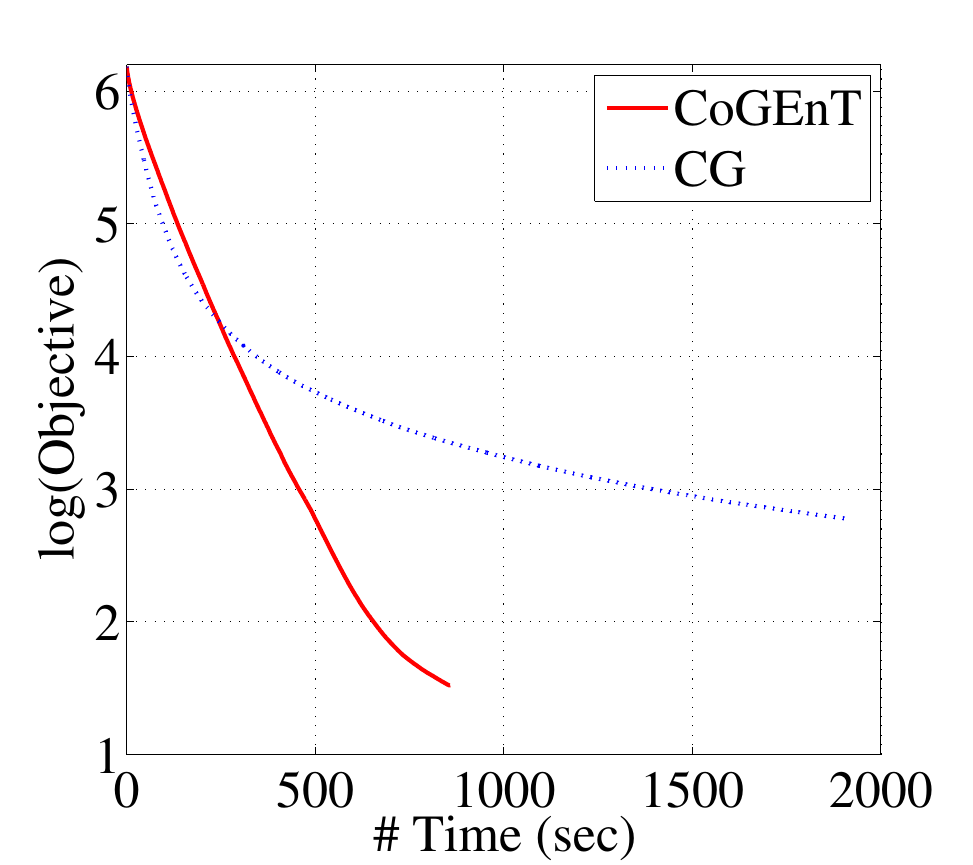}
\caption{Comparison between \cogent
  and standard conditional gradient (CG). 
}
\label{compare_iterations}
\end{figure}

To check the performance of \cogent against the the basic CG method,
we run both methods for a maximum of 5000 iterations, with a stopping
tolerance of $10^{-8}$. Fig.~\ref{compare_iterations} shows a graph of
the logarithm of the function value vs iteration count (left) and
logarithm of the function value vs wall clock time (right). On a
per-iteration basis, \cogent performs more operations than standard
CG, but the use of backward steps yields faster reduction in the
objective function value, resulting in better convergence, even when
measured in terms of run time.

To compare and contrast the effect of various steps in \cogent, we
consider a length $p = 2000$ signal and obtain $600$ Gaussian
measurements, corrupted by noise with $\sigma = 0.05$. We randomly set
$5\%$ of the coefficients of the target vector to nonzero values, and
set $\tau = \| \vx^\star \|_1$ and $\mbox{tol} = 10^{-8}$, and allow a
maximum of $1000$ iterations.  Table~\ref{tab:comparediffl1} compares
the Normalized MSE ${\| \vx^\star - \hat{\vx} \|^2}/{\| \vx^\star
  \|^2}$ and the mean $\ell_1$ error $\| \vx^\star - \hat{\vx} \|_1 /
p$ for various methods. The fully corrective variant
\cite{jaggirevisit} is only marginally better than \cogent without
truncation \footnote{Full correction was implemented using CVX
  \cite{cvx}}, but more expensive since the enhancement step only
solves the problem approximately.  Also note the significant reduction
in error once we add the truncation steps.

\begin{table}[!htp]
  \centering
\begin{tabular}{ | c |  c | c | }
\hline
\textbf{Method} & \textbf{\small NMSE $\times 100$ } & \textbf{\small L1 Error $\times 100$ } \\
\hline
$ \mbox{FW} $ & 5.848 & 0.954  \\
  \hline                       
  $\mbox{FW}_{full}$ & 2.312 &0.887 \\
  \hline
  $\mbox{CG}$ & 3.993 &0.682  \\
  \hline  
  $\mbox{CG}_{En}$ & 2.314 &0.886  \\
  \hline
    \textbf{\cogent} & $\bm{1.030}$ &$\bm{ 0.348}$   \\
  \hline
\end{tabular}
  \caption{Performance comparison for different variants of the Frank
    Wolfe (Conditional Gradient) method and \cogent. FW stands for the
    Frank Wolfe method with a step size of $\frac{2}{2+t}$,
    $\mbox{FW}_{full}$ is the fully corrective variant, CG is the
    conditional gradient method with a line search for the step size.
    $\mbox{CG}_{En}$ is \cogent without the truncation step. }
 \label{tab:comparediffl1}
  \end{table}

\begin{figure}
\centering
\includegraphics[width = 70mm, height = 35mm]{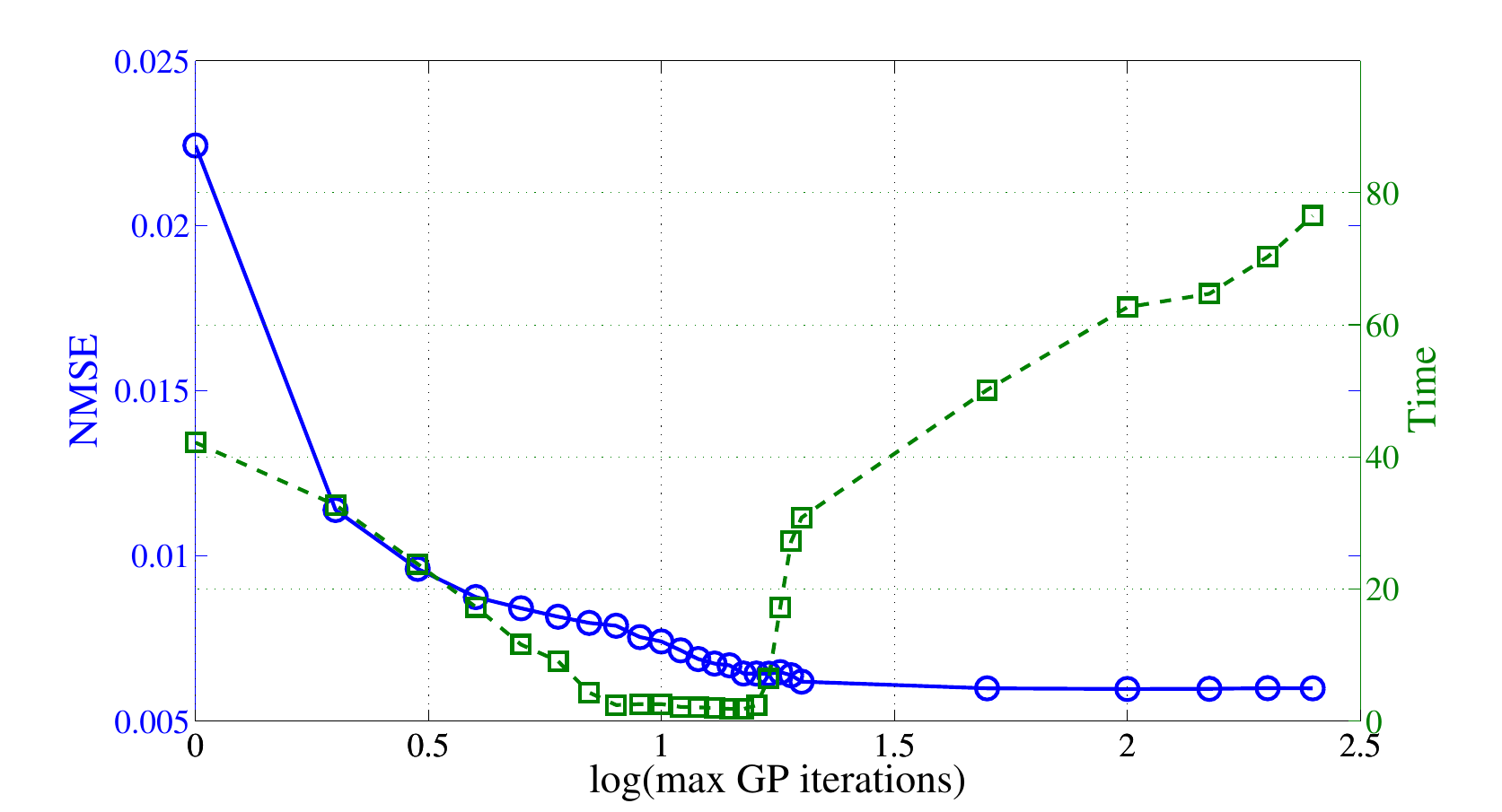}
\caption{\cogent performance as the number of steps in the inner enhancement phase is varied. }
\label{fig:enhancement}
\end{figure}

We now assess the influence of the enhancement step in \cogent. To do
so, we consider the recovery of a sparse signal of length 500, with
$10\%$ coefficients randomly set to nonzero values, with 200 noisy
($\sigma = 0.05$) Gaussian measurements. For this experiment, we chose
the convergence criterion to be the NMSE, set the maximum allowed
iterations in \cogent to be 10000, and $\tau = \| x^{\star} \|_1$. We
vary the number of projected gradient steps in the enhancement phase,
to quantify the tradeoff between the time required to solve the
enhancement subproblems and the reduction in the number of \cogent
iterations.  Fig.~\ref{fig:enhancement} plots both final NMSE and
total time taken vs the maximum number of steps allowed in the
enhancement subproblems, averaged over 15 trials. (Zero enhancement
iterations corresponds to standard CG.)  This plot shows that the best
overall solution times are obtained from a moderate-accuracy solution
of the enhancement step, with a maximum of about 10 to 15 gradient
projection iterations. (Note that we were not able to run the variants
with lower enhancement-step accuracy to full precision, as they took
too long to converge.)

We also show that \cogent is fairly robust to selection of
regularization parameter $\tau$. We recover a $100-$ sparse signal of
length $2000$ from $500$ Gaussian measurements, corrupted by AWGN
$\sigma = 0.05$.  We varied $\tau$ by various powers of $2$ about its
optimal value $\tau^*$, which is the $\ell_1$ norm of the true
signal. Note that the NMSE does not change dramatically, and is in
fact quite insensitive to $\tau$ when $\tau$ is greater than $\tau^*$.

Table~\ref{tau_effect} shows that the NMSE does not
drastically increase, even when $\tau$ is made very large. We varied
$\tau$ as $\tau^\star \times 2^{t}$, for different $t$, $\tau^\star$
being the $\ell_1$ norm of the true signal.

\begin{table}[!htp]
  \centering
   \resizebox{\linewidth}{!}{
\begin{tabular}{ | c |  c | c | c |  c | c |  c |  c | }
\hline
\textbf{t} & -2 & -1 & 0 & 1 & 2 & 3 & 4 \\
\hline
\textbf{NMSE} & 0.172 & 0.131 & 0.039 & 0.057 & 0.058 & 0.059 & 0.063 \\
  \hline
\end{tabular}}
  \caption{NMSE for sparse recovery with $\tau= 2^t\tau^*$ for various
    $t$.}
 \label{tau_effect}
  \end{table}


We next compare \cogent with some other \emph{greedy} methods for
solving $\ell_1$ norm constrained problems: CoSaMP \cite{cosamp},
Subspace Pursuit \cite{sp}, and GraDes \cite{grades}\footnote{All
  codes were obtained directly from the Internet, and were used
  without modification}. We consider a sparse vector of length $p =
20000$ with $s=1000$ nonzeros randomly chosen, with $n = 5000$ noisy
Gaussian measurements, with $\sigma = 0.05$. All codes were run for a
maximum of $1000$ iterations, with a convergence tolerance of
$10^{-4}$.  Results are averaged over 5 independent
trials. Table~\ref{tab:time_l1} shows that GraDeS is the fastest, but
with a poor solution quality. CG is fast, but with a lack of
enhancement and truncation steps, the NMSE is still an order of
magnitude worse than \cogent and SP.  Only SP has an NMSE comparable
to that of \cogent, but is much slower. Again, we set $\tau = \| \vx
\|_1$ for \cogent and CG, while the other algorithms require the true
signal sparsity, which we supply as a parameter.
 
\begin{table}[!htp]
  \centering
\begin{tabular}{ | c |  c | c | }
\hline
\textbf{Method} & \textbf{\small Time (seconds) } & \textbf{\small NMSE } \\
\hline
$ \mbox{CoSaMP}$	& $1479.9$	& 1.5584  \\
  \hline                       
  $\mbox{SP}$		& $3601.1 $ 	& 0.0488 \\
  \hline
  $\mbox{CG}$ 		& $456.96$ 	&0.2185  \\
  \hline  
  $\mbox{GraDes}$ 	& $\bm{405.80}$  	& 1.0605  \\
  \hline
    \cogent 			&  $1041.6$ 	& $\bm{0.0436}$   \\
  \hline
\end{tabular}
  \caption{Timing comparisons for CG, \cogent, Subspace Pursuit (SP), CoSaMP and GraDes.  }
 \label{tab:time_l1}
  \end{table}


\begin{figure}[!h]
\centering 
\includegraphics[width = 60mm, height = 35mm]{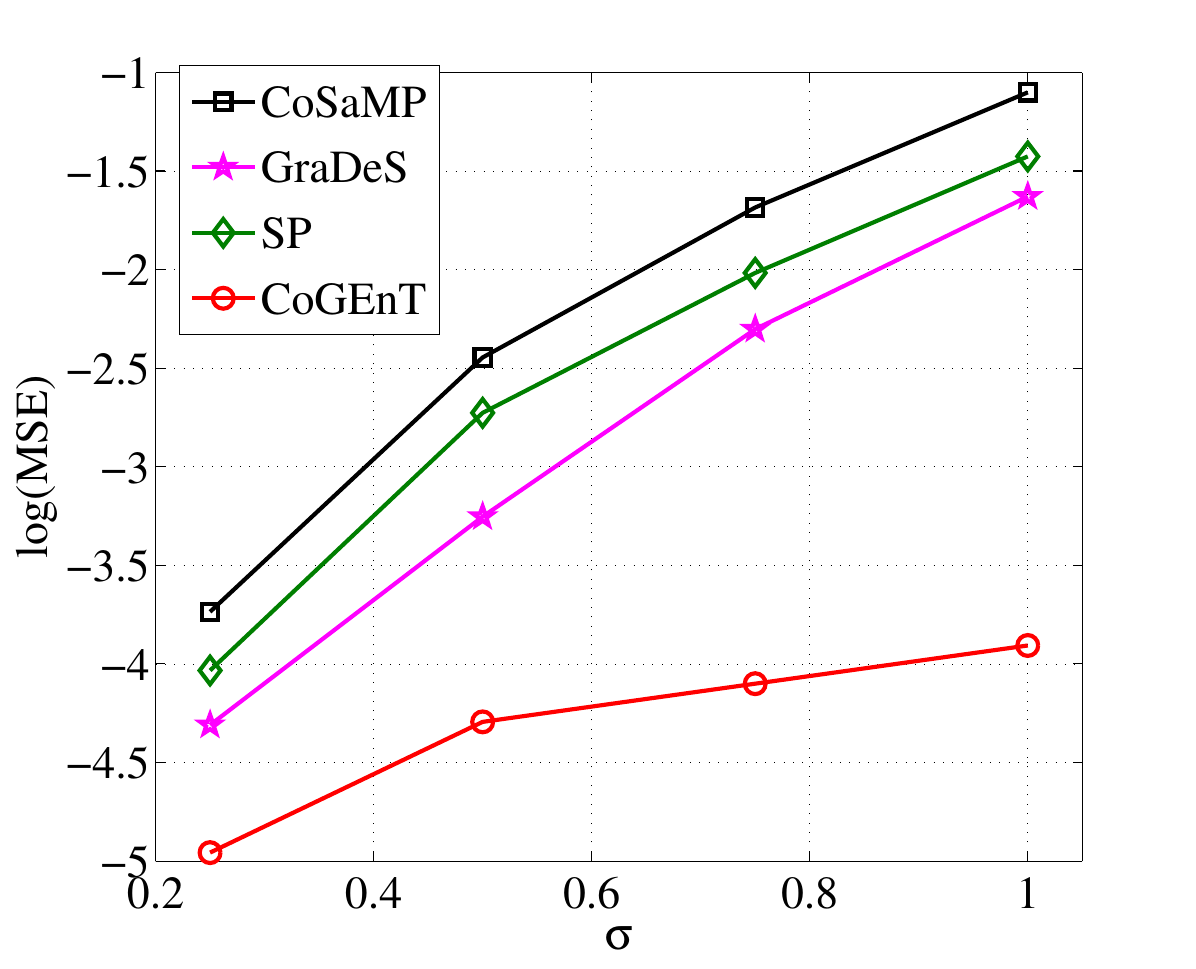}
\caption{Comparison of solution quality obtained by different methods. }
\label{comparemany}
\end{figure}
Fig.~\ref{comparemany} shows a comparison of solution quality among
the same methods. As a performance metric, we used the mean square
error between the true and predicted vectors. We performed 10
independent trials, setting $\mPhi$ in each trial to be a $3000 \times
10000$ matrix, with reference solution $\vx^\star$ chosen to have
$s=300$ nonzeros, randomly assigned. Observations $\vy$ were corrupted
with AWGN with standard deviation $\sigma$ in the range $[0,1]$. For
\cogent and CG, we chose $\tau:= \| \vx^\star \|_1$. For CoSaMP,
GraDeS and Subspace Pursuit methods, we set $s=300$, the known
sparsity level of the optimal signal.  \cogent performs better at all
levels of noise, we believe due to the flexibility of the
atomic-constrained formulation, as opposed to a hard limit on the
number of basis elements.



\vspace{-3mm}
\subsection{Overlapping Group Lasso}
In group-sparse variants of \eqref{eq:cs} we seek vectors $\vx$ such
that $\mPhi \vx \approx \vy$ for given $\mPhi$ and $\vy$, such that
the support of $\vx$ consists of a small number of predefined groups
of the coefficients. We denote each group by $G \subset
\{1,2,\dotsc,p\}$ and denote the full collection of groups by $\G$.
CG and \cogent do not require replication of variables, as is done in
prox-linear algorithms \cite{jacob,nricip11}. The atom selection step
(Step~\ref{greedy} in Algorithm~\ref{alg:cogent}) amounts to the
following operation:
\begin{align*}
\hat{G} &=\arg \max_{G \in \G} \| [\nabla(f(\vx_t))]_{G} \|, \\ 
[\va_{t+1}]_{\hat{G}} &= -[\nabla f(\vx_t))]_{\hat{G}} / \| [\nabla f(\vx_t))]_{\hat{G}} \|, \\
 \ [\va_{t+1}]_{i} &= 0 \;\; \mbox{for $i \notin \hat{G}$.}
\end{align*}

%

We compare the performance of \cogent with an accelerated proximal
point (PP) approach \cite{sparsa} that uses variable replication.
We considered $M$ group sparse signals with $\floor{M/10}$ groups
chosen to be active in the reference solution, where each group has
size 50. The groups are ordered in linear fashion with the last 30
indices of each group overlapping with the first 30 of the next
group. We then took $n = \ceil{p/2}$ measurements with a Gaussian
sensing matrix $\mPhi$, with AWGN of standard deviation $\sigma=0.1$
added to the observations. Table~\ref{timelgl} shows runtimes for the
two approaches. We set the maximum iterations to be $2n$ and the
tolerance to be $10^{-5}$. We searched for $\tau$ (\cogent) and
$\lambda$ (prox) over a grid, and choose the values that yield best
MSE performance.

\begin{table}[!htp]
  \centering
\begin{tabular}{ | c |  c | c | r | r |}
\hline
\textbf{M} & \textbf{\small True } & \textbf{\small Replicated } & \textbf{\small time }  & \textbf{\small time }\\
\textbf{} & \textbf{\small  Dimension} & \textbf{\small  Dimension} & \textbf{\small  \cogent}  & \textbf{\small  PP}\\
\hline
100 & 2030 & 5000 & 15 & 22 \\
  \hline                       
  1000 & 20030 &50000 & $211$  & $462 $\\
  \hline
  1200 & 24030 &60000 & $359$ &  $778$ \\
  \hline  
  1500 & 30030 &75000 & $575$  & $1377$  \\
  \hline
    2000 & 40030 &100000 & $852$  & $2977$  \\
  \hline
\end{tabular}
  \caption{Recovery times (in seconds) for \cogent and prox-linear methods applied to a synthetic  overlapping group-sparse problem.}
 \label{timelgl}
  \end{table}

\vspace{-3mm}
\subsection{Matrix Completion}

In low-rank matrix completion, the atoms are rank-one matrices and the
observations are individual elements of the matrix.  If $(\vu, \vv)$
are the first left and right singular vectors of $- \nabla f_t$, the
solution of Step~\ref{greedy} in Algorithm~\ref{alg:cogent} is $
\va_{t+1} = \vu \vv^T$. The cost of finding only the top singular
vectors in the gradient matrix is smaller than the cost of a full SVD
by a factor about equal to the smaller dimension of the matrix.

\begin{figure}
\centering
\includegraphics[width = 60mm, height = 40mm]{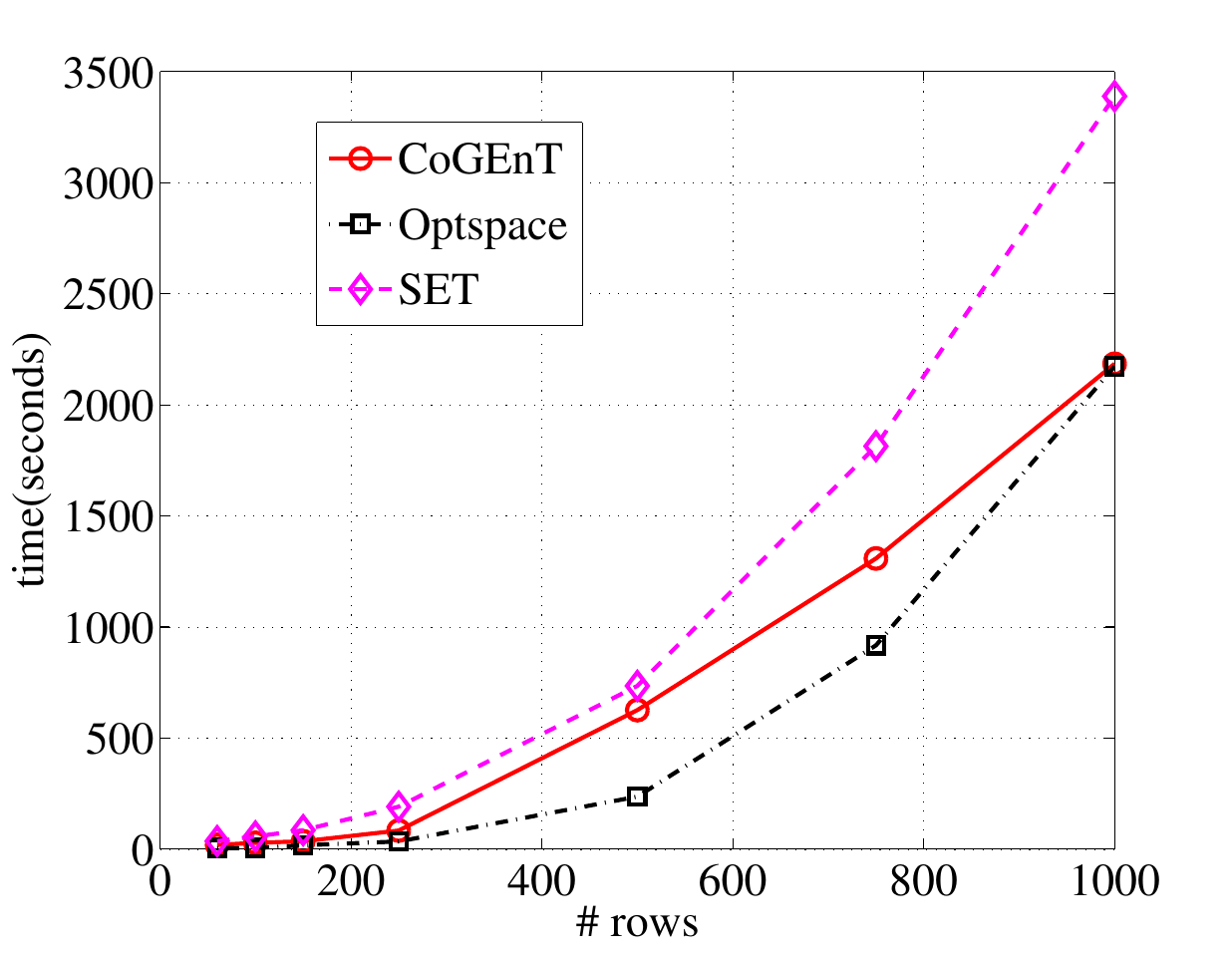}
\caption{Speed comparison for Matrix Completion. Results are averaged over 10 trials}
\label{matcomp_times}
\end{figure}

We compared \cogent with Optspace \cite{optspace} and SET \cite{set},
on matrices of dimension $m \times \ceil{\frac{4m}{3}}$ with rank
$\max\{3,\ceil{\frac{n}{100}}\}$. The matrices were randomly generated
as $UV^T$, with $U$ and $V$ obtained by orthogonalizing random
Gaussian matrices of appropriate size. We sampled $20\%$ of the matrix
elements and ran all methods until convergence (with a maximum of 5000
iterations and tolerance $10^{-4}$). From Fig.~\ref{matcomp_times}, we
see that \cogent is faster than SET as the matrix size increases. In
fact, SET does not scale well for larger sizes, a regime where \cogent
is still a viable option because of its lower computational
cost. Optspace is typically faster, but since it is an alternating
minimization method, it may be attracted to local minima. (Recent
results have shown that alternating minimization approaches do
converge to the global optimum, provided they are initialized
appropriately \cite{jain_altmin}.) In terms of reconstruction error,
Optspace typically yielded slightly higher error rates compared to SET
and \cogent (see Table~\ref{tab:materr}).

\begin{table}[!htp]
  \centering
\begin{tabular}{ | c |  c | c | c | }
\hline
\textbf{$\#$ rows} & \textbf{\small SET } & \textbf{\small Optspace } & \textbf{\small \cogent } \\
\hline
60 & 0.3529 & 0.4828 & $\bm{0.3518}$ \\
\hline
100 & 0.1935 & 0.2969 & $\bm{0.1933}$ \\
\hline
150 & 0.1524 & 0.1821 & $\bm{0.1506}$ \\
\hline
250 & 0.0824&  0.1214 & $\bm{0.0817} $\\
\hline
500 & 0.0367  & 0.0493 &$\bm{ 0.0363}$\\
\hline
750 & 0.0247 & 0.0282 & $\bm{ 0.0244}$ \\
\hline
1000 & $\bm{0.0184}$ & 0.0310 &$\bm{ 0.0184}$ \\
\hline
\end{tabular}
  \caption{MSE $\times 1000$ for the Matrix Completion methods
    considered. Note that alternating minimization (Optspace) yields
    higher errors than the other methods.}
 \label{tab:materr}
  \end{table}

\section{Experiments: Novel Applications}
\label{sec:novelapps}

We now report on the application of \cogent to recovery problems in
several novel areas of application. In some cases, \cogent and CG are
the only practical approaches for solving these problems, while in
others, the fact that the signal to be recovered has an ``atomic"
representation allows \cogent to be used for recovery.

\subsection{Tensor Completion}

Recovery of low-rank tensor approximations arises in applications
ranging from multidimensional signal processing to latent-factor
models in machine learning \cite{animatensor}.  Here, we consider the
recovery of symmetric orthogonal tensors from incomplete measurements
using \cogent. We seek a tensor $T$ of the form $T = \sum_{i = 1}^r
c_i [\otimes \vu_i]$, where $\otimes \vu$ indicates an $t$-fold tensor
product of a vector $\vu \in \R^p$ such that $\| \vu \|=1$. We obtain
partial measurements of this tensor of the form $y=\mathcal{M}\left( T
\right)$, where $\mathcal{M}\left( \cdot \right)$ is a \emph{masking
  operator} that reveals a certain subset of the entries of the
tensor. We formulate this problem in an atomic norm setup, wherein the
objective function (which captures fidelity to the observations) is
$f(T):=\frac{1}{2} \|y-\mathcal{M}\left(T \right) \|^2$. The atomic
set has the form
\[
\A = \{  \otimes \vu \, : \, \vu \in \R^p, \; \|\vu\|_2=1 \}.
\]
In applying \cogent to this problem,
the greedy step requires calculation of the symmetric rank-one tensor
that best approximates the gradient of the loss function. This
calculation can be performed efficiently using power iterations
\cite{animatensor}.
We implement a backward step based on basis reoptimization and
thresholding (Algorithm 3), where the new basis is obtained from a
tensor decomposition, computed via power iterations.

{

Next, we describe and interpret our numerical experiments. We consider
randomly generated orthogonally decomposable tensors of order $3$ and
rank $r=3$ of the form $\vx^* = \sum_{i=1}^{3} c_i v_i^{\otimes}$ in
all our experiments. The components $v_i$ are chosen by picking random
orthonormal sets of vectors in $\R^{n}$. In our experiments we vary
the dimension of the tensor from $n=20$ to $n=60$. Each entry of the
tensor is revealed with a specified probability $p$. We vary the
sampling fraction defined as the fraction ${m}/{n^3}$, where $m$ is
the (expected) number of revealed entries. We conduct a number of
trials for each pair of tensor size ($n$) and sampling fraction. In
each trial, we declare recovery to be successful if the difference
between the true tensor and the recovered tensor (in Euclidean norm)
is less than $10^{-4}$.

Our experiments compare two approaches. The first approach involves a
matricization approach described in
\cite{squaredeal,gandy2011tensor,liu2013tensor}, solved using standard
a matrix completion approach, implemented in Matlab. The second
approach uses \cogent, making use of backward steps, with forward
steps computed using power iterations. The parameter $\tau$ as always
is set to be the $\ell_1$ norm of the true coefficients of the
solution.

In Fig.~\ref{fig:prob_rec}, we fix the tensor size at $n=20$ and
plot the empirical probability of success as a function of the
sampling fraction, using $20$ trials per choice of sampling
fraction. Our tensor atomic-norm approach substantially outperforms
the matrix unfolding approach. In Fig.~\ref{fig:phase} we plot the
phase transition plots for the two approaches. The $x$ axis we shows
the sampling fraction, while the $y$ axis shows the value of $n$.  For
each coordinate square, we conduct $10$ trials, and plot the empirical
probability of success in grayscale, with white representing $1$ and
black representing $0$. We observe that the atomic-norm approach
substantially outperforms the matrix unfolding based approach. For
instance, for $n=50$, the matricization approach is unable to reliably
recover tensors with sampling fractions below $0.45$, whereas the
atomic norm approach reliably recovers the same even at sampling
fractions of $0.1$ (the lowest tried in this set of experiments). It
seems that the atomic-norm formulation is more powerful than the
matricization approach, and that \cogent is effective in solving it.

        \begin{figure}
 \centering
         \includegraphics[scale=0.25]{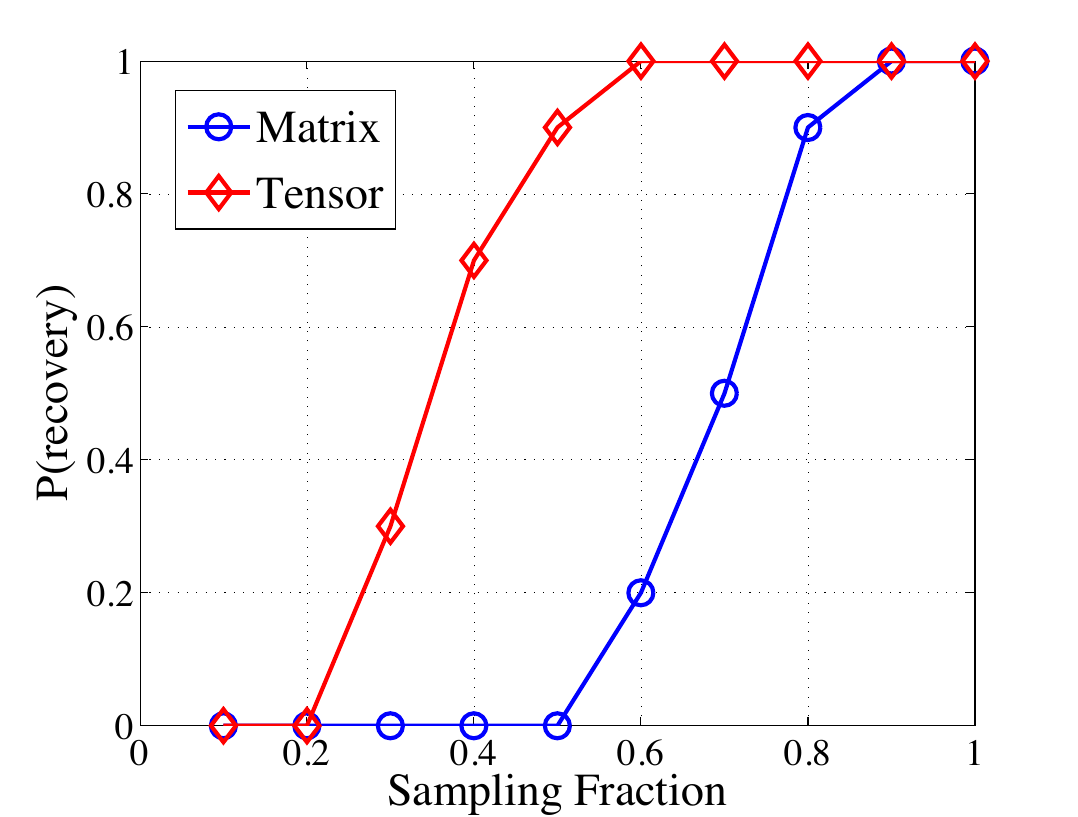}
\caption{Comparison of success probabilities for tensors of size $n=20$. 
}
                \label{fig:prob_rec}
        \end{figure}

\begin{figure}
        \centering
\includegraphics[width=.24\textwidth]{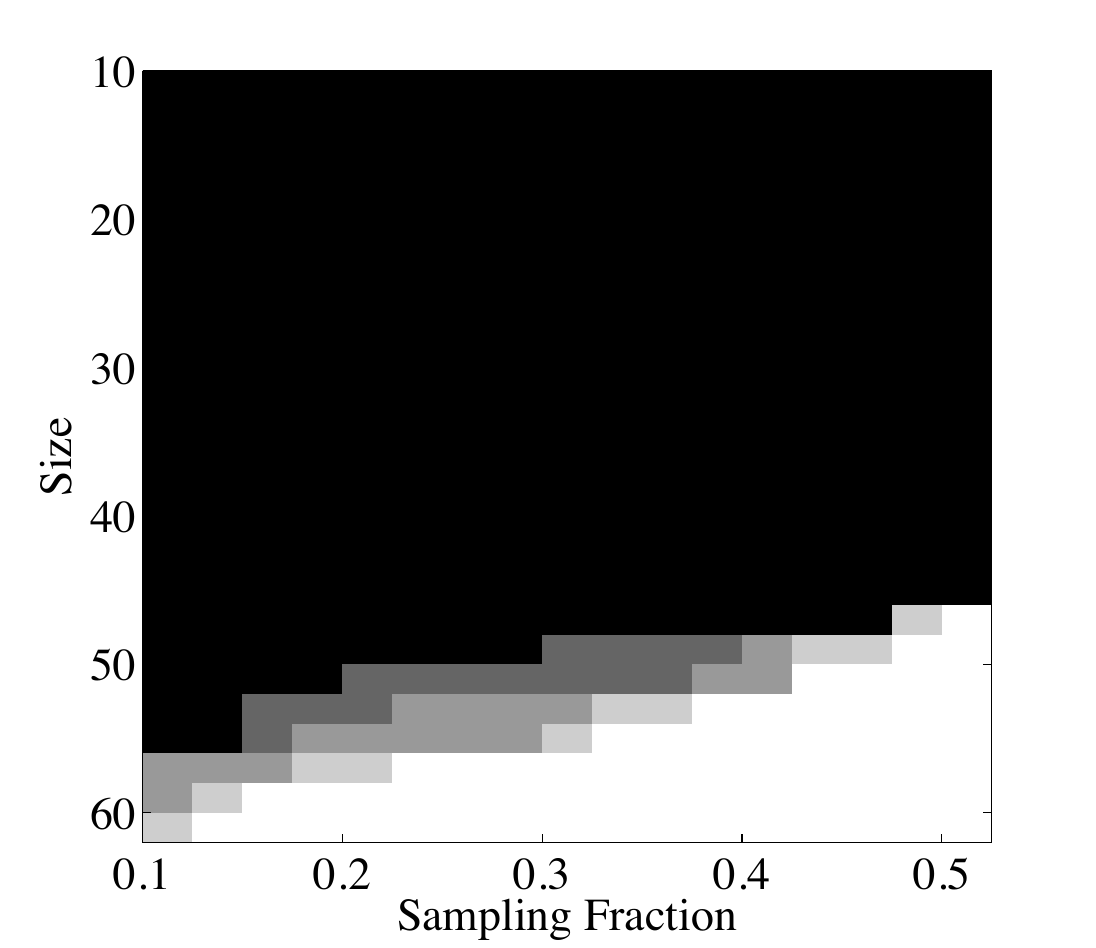}
\includegraphics[width=.24\textwidth]{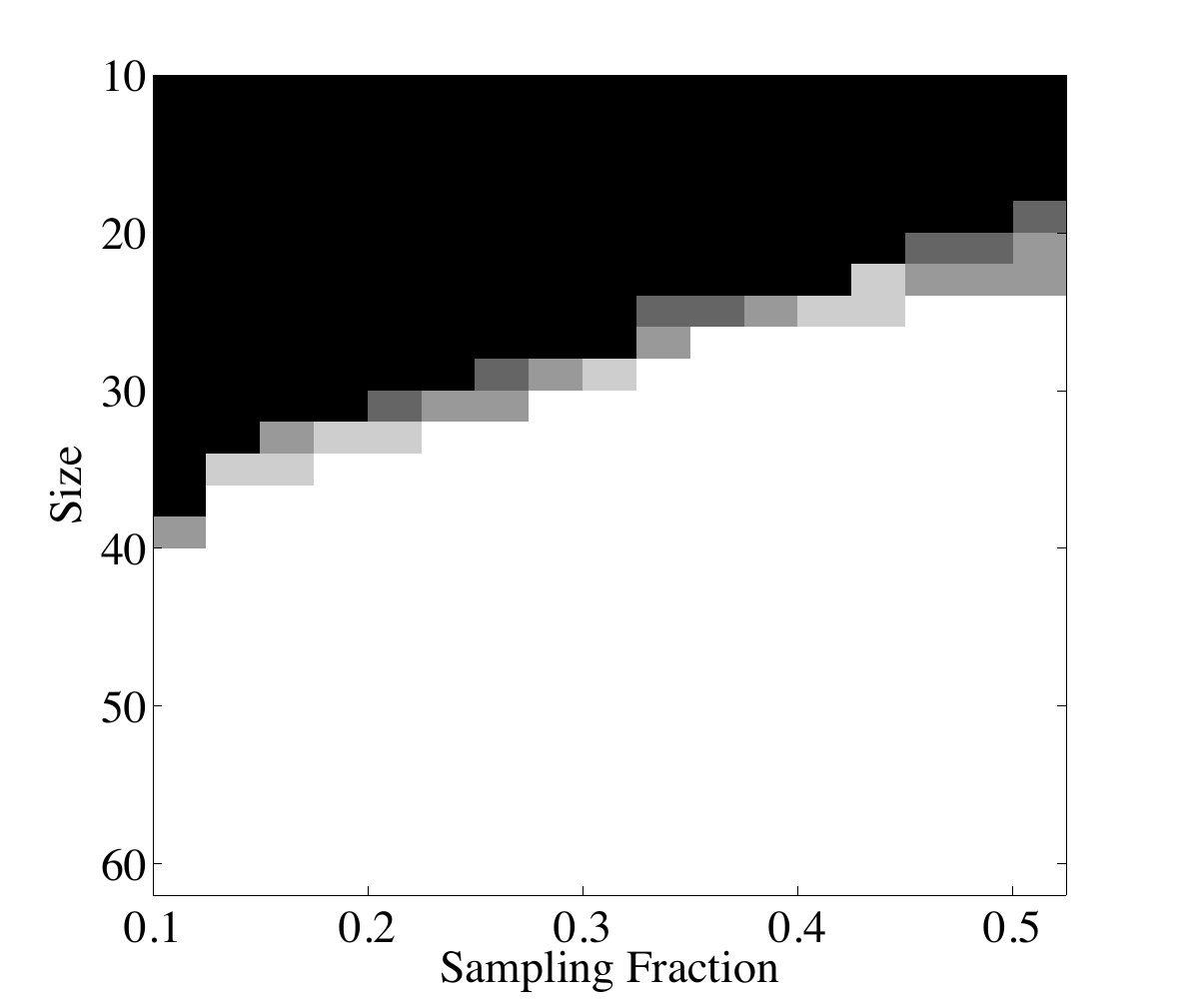}
                \caption{Phase transition plots based on matricized nuclear norm (left) and atomic norm (right) minimization. White indicates success and black indicates failure in recovery. }
                \label{fig:phase}
        \end{figure}


\color{black}

\subsection{Moment Problems in Signal Processing} \label{sec:moments}

Consider a continuous time signal 
\[
\phi(t)=\sum_{j=1}^{k} c_j \exp(i 2 \pi f_j t), 
\]
for frequencies $f_j \in [0,1]$, $j=1,2,\dotsc,k$ and coefficients
$c_j>0$, $j=1,2,\dotsc,k$. In many applications of interest, $\phi(t)$
is sampled at times $S:=\left\{t_i\right\}_{i=1}^n$ giving an
observation vector $\vx := [\phi(t_1), \phi(t_2), \ldots, \phi(t_n)]
\in \mathbb{C}^n$. The observed information is therefore
\[
\vx =\sum_{j=1}^{k} c_j a(f_j),
\] 
where
\[
 a(f_j) =\left[ e^{i 2\pi f_j t_1}, e^{i 2\pi f_j t_2}, \; \dotsc,
   e^{i 2\pi f_j t_n} \right]^{T}.
\]
Finding the unknown coefficients $c_j$ and frequencies $f_j$ from
$\vx$ is a challenging problem in general. A natural convex
relaxation, analyzed in \cite{offgrid}, is obtained by setting
$\mPhi=I$ in $\eqref{eq:opt1}$ and defining the atoms to be $a(f)$ for
$f \in [0,1]$, a set of infinite cardinality.

The main technical issue in applying \cogent to this problem is the
greedy atom selection step (Step~\ref{greedy} of
Algorithm~\ref{alg:cogent}), which requires us to find the maximum
modulus of a trigonometric polynomial on the unit circle. This
operation can be formulated as a semidefinite program
\cite{bounded_real_lemma}, but since SDPs do not scale well to high
dimensions \cite{offgrid}, this approach has limited appeal.
In our implementation of \cogent, we form a discrete grid of frequency
values, starting with an initial grid of equally spaced frequencies,
refined between iterations by adding new frequencies midway between
each pair of selected frequencies. This approach differs from
\cite{offgrid} in that although the initialization is via a grid of
frequencies, the ability to refine the grid on a per-iteration basis
allows us to achieve much higher precision using fewer grid
points. The presence of the backward step in \cogent means that the
method can discard coarse grid points previously selected in favor of
better grid points added in later iterations. This also means that the
initialization can be quite coarse, another advantage over
\cite{offgrid}.

%
By controlling the discretization in this way, we are essentially
controlling and refining the inexactness of the forward step (as captured by \eqref{approxgreedy}).  Indeed, the accuracy
required in \eqref{approxgreedy} can provide guidance for the adaptive
discretization process. Step~\ref{greedy} simply selects an atom
$a(f)$ corresponding to the frequency $f$ in the current grid that
forms the most negative inner product with the gradient of the loss
function.



Our implementation of the backward step for this problem has two
parts. Besides performing Algorithm~\ref{alg:bs_multiple} to remove
multiple uninteresting frequencies, we include a heuristic for merging
nearby frequencies, replacing multiple adjacent spikes by a single
spike, when it does not degrade the objective too much to do so. Fig.~\ref{fig:freq} compares the performance of \cogent with that of
standard CG on a signal with ten uniformly randomly chosen frequencies in
$[0,1]$. We take samples at $300$ timepoints of a signal of length
1000, corrupted with AWGN with standard deviation $.01$.  The left
figure in Fig.~\ref{fig:freq} shows the signal recovered by \cogent,
indicating that all but the smallest of the ten spikes were recovered
accurately. The critical role played by the backward step can be seen
by contrasting these results with those reported for CG in the right
figure of Fig.~\ref{fig:freq}, where many spurious frequencies
appear.

\begin{figure}[!h]
\centering
  \includegraphics[width=.24\textwidth]{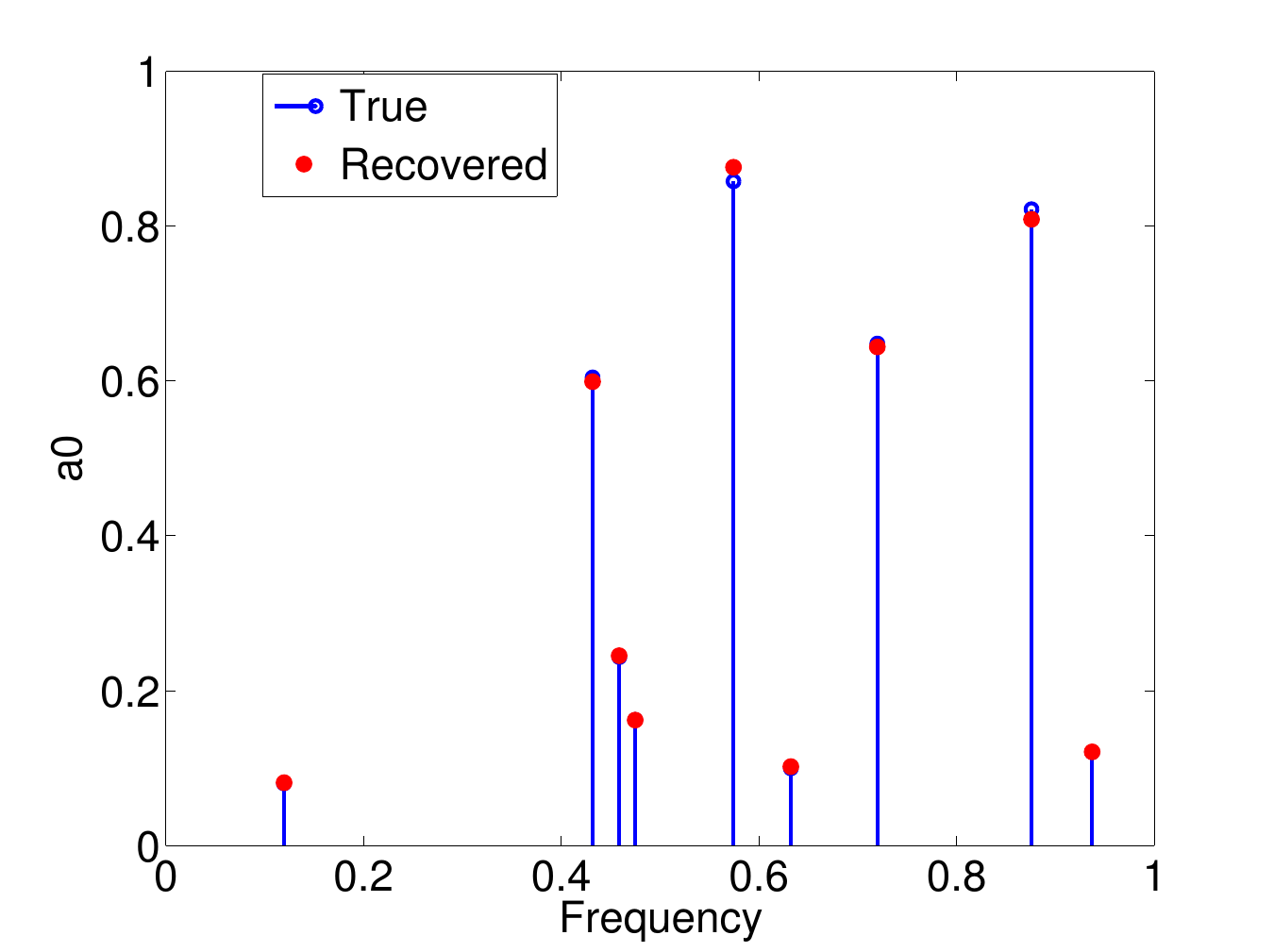}
\includegraphics[width=.24\textwidth]{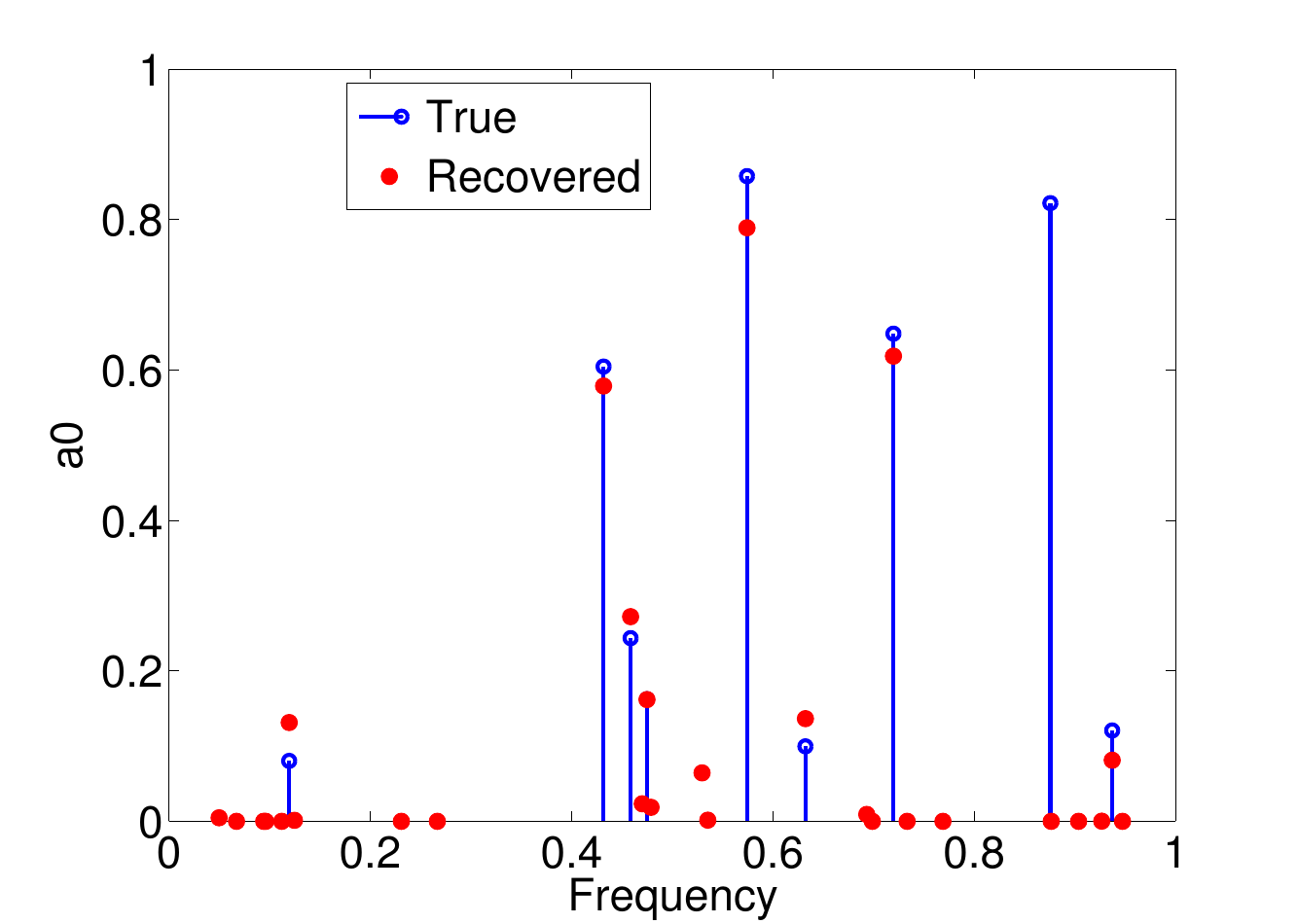}
\caption{\cogent and CG for  off-grid compressed sensing. Blue
  spikes and circles represent the reference solution, and red circles
  are those estimated by the algorithms.}
\label{fig:freq}
\end{figure}
%

We compared \cogent to the SDP formulation as explained in
\cite{offgrid}. Although the SDP solves the problem exactly, it does
not scale well to large dimensions, as we show in the timing
comparisons of Fig.~\ref{offgrid_speed}. The instances were generated
as follows: For a particular signal length $p$, we randomly choose 5
frequencies to be active, choosing values from a uniform $(0,1]$
distribution. We obtain ${p}/{4}$ noisy measurements with $\sigma =
0.1$. We run \cogent with $2n$ iterations and a tolerance of
$10^{-10}$.

\begin{figure}
{\centering
\includegraphics[width=.25\textwidth]{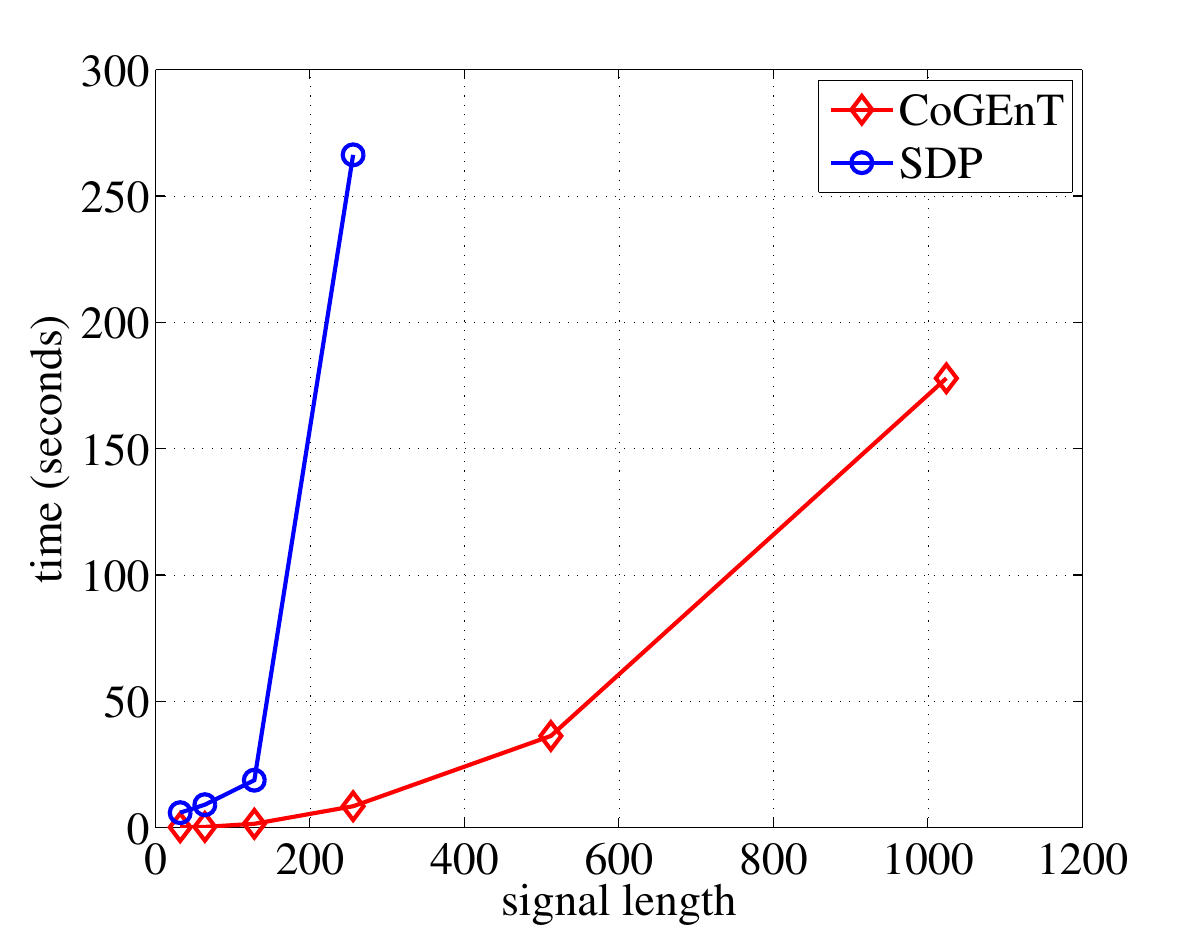}
\caption{Speed comparison with SDP.  The SDP formulation does not
  scale well.}
\label{offgrid_speed}
}
\end{figure}

The formulation above can be generalized to include signals that are a
conic combination of a few arbitrary functions of the form
$\phi(t,\alpha_i)$.
%
\begin{itemize}
\item Bessel and Airy functions form natural signal ensembles that
  arise as solutions to differential equations in physics. As an
  example, letting $J_r(t)$ denoting Bessel functions of the first
  kind, we have
\[
\phi(t;\alpha_1, \alpha_2, \alpha_3)=J_{\alpha_{1}}\left( \frac{t}{\alpha_2}-\alpha_3 \right),
\]
where $\alpha_1, \alpha_2, \alpha_3 \in \mathbb{R}_{+}$.  Here, each
atom is defined by a specific choice of the triple
$(\alpha_1,\alpha_2,\alpha_3)$, leading to an atomic set $\A$ with
infinite cardinality.
\item Triangle and sawtooth waves. Consider for instance the sawtooth
  functions:
\[
\phi(t;\alpha_1, \alpha_2)=\frac{t}{\alpha_1}-\floor*{\frac{t}{\alpha_1}}-\alpha_2,
\]
where $\alpha_1, \alpha_2 \in \mathbb{R}_{+}$. Each atom is defined by
a specific choice of $(\alpha_1,\alpha_2)$.
Fig.~\ref{fig:sawtooth} shows successful
recovery of a superposition of sawtooth functions from a limited
number of samples.



\begin{figure}
\centering \subfloat[The true signal (blue) is a superposition of
  sawtooth functions. Red dots show samples acquired.]{
  \centering
  \includegraphics[width=.2\textwidth]{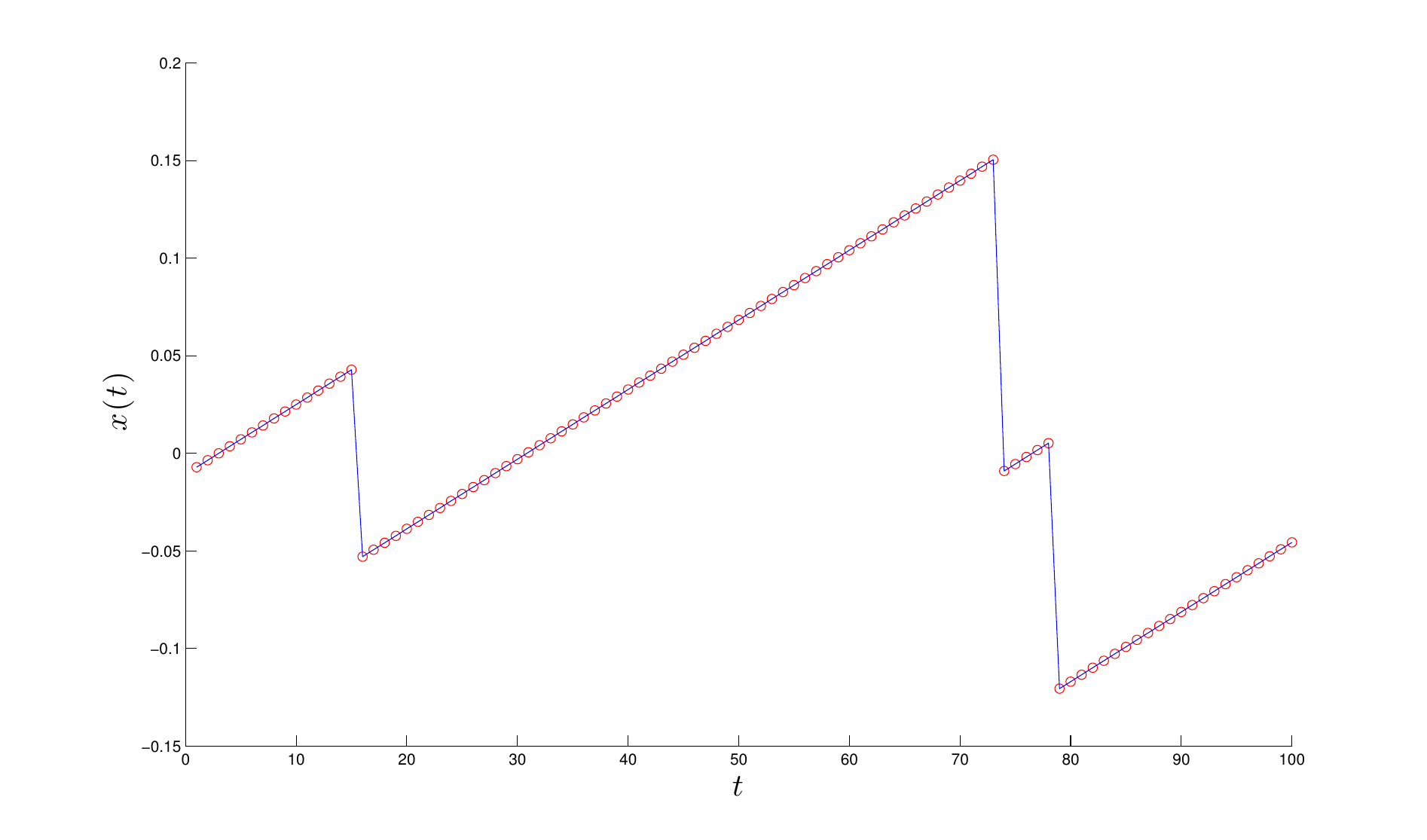}
\label{graph_group_setup}
} \quad \subfloat[Sawtooth components recovered by \cogent.]{ \centering
  \includegraphics[width=.2\textwidth]{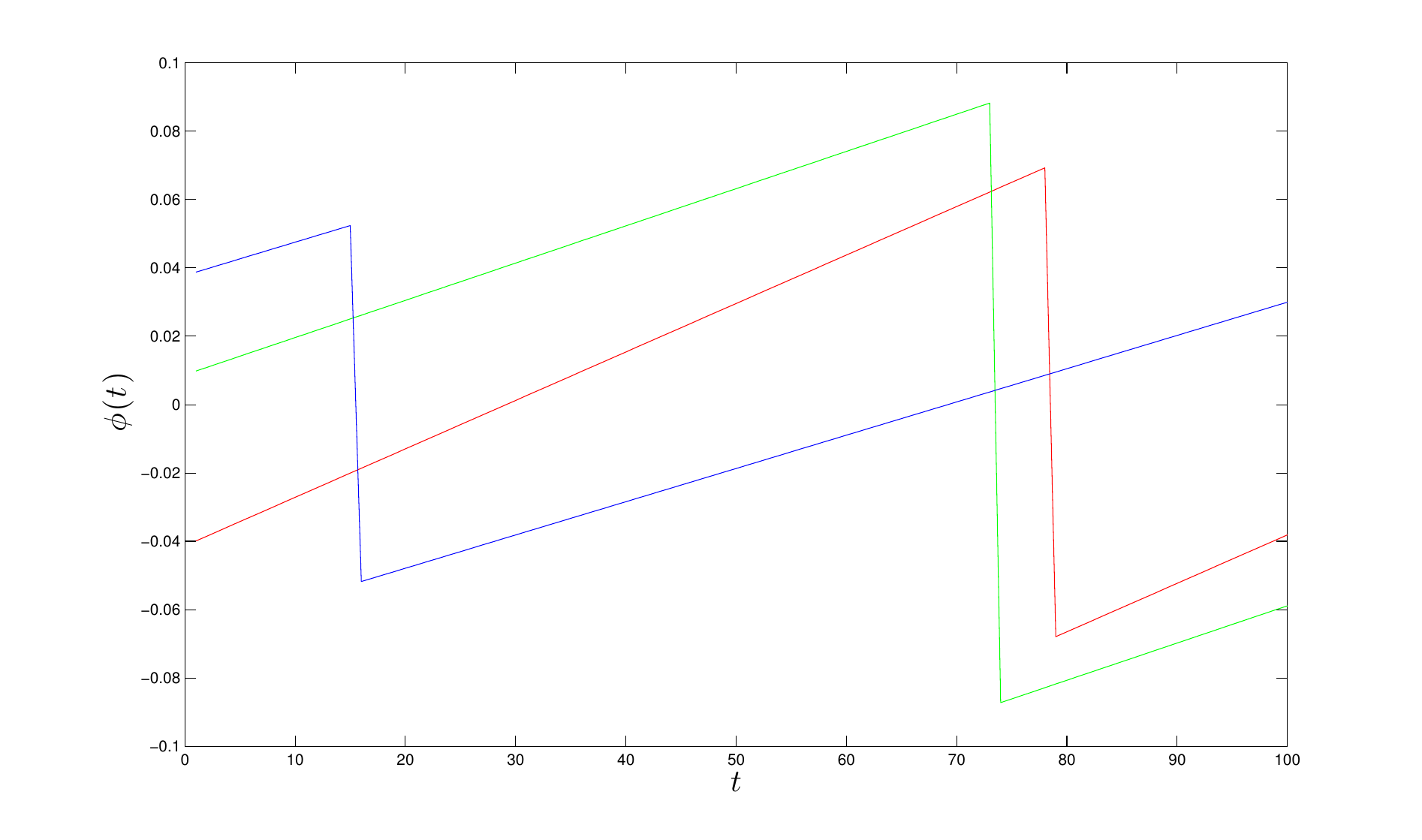}
\label{fig:sawtooth_b}
}
\caption{Recovering  sawtooth components by sampling. (Best seen in color)}
\label{fig:sawtooth}
\end{figure}

\item Ricker wavelets arise in seismology applications, with the atoms
characterized by $\sigma>0$:
\[
\phi(t;\sigma)=\frac{2}{\sqrt{3 \sigma} \pi^{\frac{1}{4}}}\left(1-\frac{t^2}{\sigma^2} \right) \exp\left(-\frac{t^2}{2 \sigma^2} \right).
\]
\item Gaussians, characterized by parameters $\mu$ and 
  $\sigma$:
\[
\phi(t;\mu, \sigma)=\frac{1}{\sqrt{2 \pi} \sigma} \exp\left( -\frac{(t-\mu)^2}{2 \sigma^2}\right).
\]
Estimating Gaussian mixtures from sampled data is a much-studied
problem in machine learning.
\end{itemize}

The key ingredient in solving these problems within the atomic norm
framework is efficient (approximate) solution of the atom selection
step. In some cases, this can be done in closed form, whereas for all
the signals mentioned above, approximate solutions can be obtained via
adaptive discretization.

\subsection{OSCAR}


The regularizer for the Octagonal Shrinkage and Clustering Algorithm
for Regression (OSCAR) method is defined for $\vx \in \R^p$ as
follows:
\[
\| \vx \|_1 + c \sum_{j = 1}^p \sum_{k = 1}^j \max{\{ |\vx_j|, |\vx_k| \}}
\]
In \cite{mario_oscar}, the authors show that this indeed can be
expressed as an atomic norm, and also give an efficient method to find
the next atom to add in the forward greedy step.  We compared \cogent
for OSCAR with the proximal point based scheme \cite{candes_sortedl1},
of which OSCAR is a special case. (For a comparison of different
prox-based methods, we refer to the interested reader to
\cite{oscar_prox}.) We considered a length $5000$ vector $x$ of the
following form:
\[ a = \theta_a + \zeta_a  ,~\ b = \theta_b + \zeta_b, ~\ c = \theta_c + \zeta_c ~\ d = \theta_d + \zeta_d
\]
where the $\zeta$ are vectors of length $20$ with $i.i.d.$ Gaussian
entries, and $\theta \sim \mathcal{U}[-1,1]$. In MATLAB notation,
$x(1:20) = a, ~\ x(301:320) = b, ~\ x(801:820) = c, ~\ x(1001:1020) =
d$. We obtained 500 Gaussian measurements and corrupted them with
varying amounts of noise $\sigma$. We allowed both methods to run for
at most $1000$ iterations, with a convergence tolerance of
$10^{-6}$. Fig.~\ref{fig:oscar} shows the results we obtained, in
terms of MSE ${\| \hat{\vx} - \vx \|^2_2}/{1000}$. We do not report
timing results in this case, since we were comparing our MATLAB code
with C code. However, the standard CG method has been observed to be
faster than prox-based methods, so it is reasonable to expect \cogent
to be faster as well, when implemented appropriately.


\begin{figure}
\centering
\includegraphics[width = 55mm, height = 35mm]{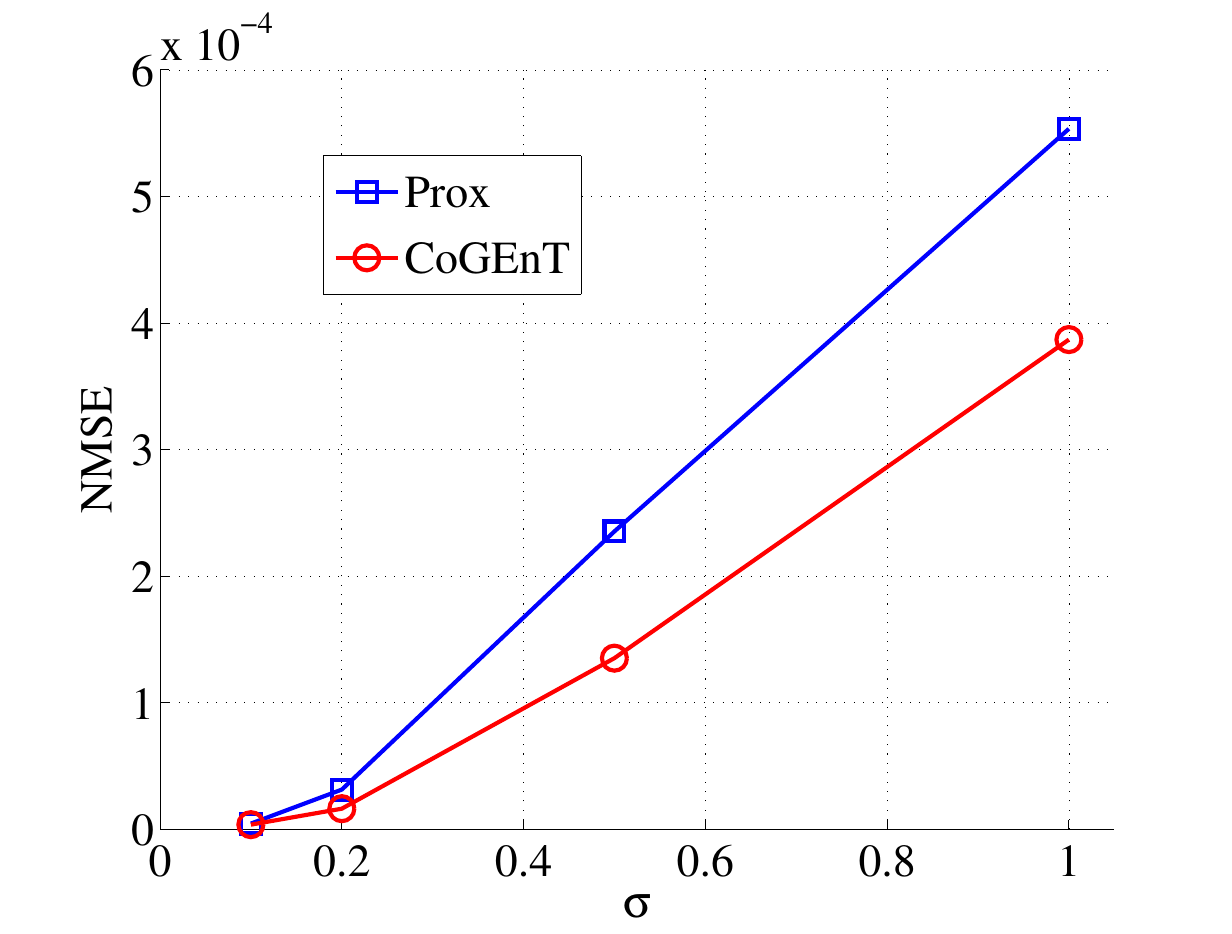}
\caption{Comparison between \cogent and proximal-point methods for
  solving the OSCAR problem. We found \cogent to be more robust to
  noise.}
\label{fig:oscar}
\end{figure}

\section{Reconstruction and Deconvolution} 
\label{sec:decon}

The deconvolution problem involves recovering a signal of the form
$\vx =\vx^1+\vx^2$ from observations $\vy$ via a sensing matrix
$\mPhi$, where $\vx^1$ and $\vx^2$ can be expressed compactly with
respect to different atomic sets $\A_1$ and $\A_2$.  
We mentioned several instances of such problems in
Section~\ref{sec:intro}.
Adopting the optimization-driven approach outlined in
Section~\ref{sec:intro}, we arrive at the following  convex
optimization formulation:
\begin{align*}
\underset{\vx^1, \vx^2}{\text{minimize}} & \qquad \frac12 \|\vy-\mPhi(\vx^1+\vx^2)\|^2 \\
\text{subject to} & \qquad \|\vx^1 \|_{\A_1} \leq \tau_1 ~\mbox{and} ~\|\vx^2 \|_{\A_2} \leq \tau_2.
\end{align*}

Algorithm \ref{alg:cogent} can be extended to this situation, as we
describe informally now.
Each iteration starts by choosing an atom from $\A_1$ that nearly
minimizes its inner product with the gradient of the objective
function with respect to $\vx^1$; this is the forward step with
respect to $\A_1$. One then performs a backward step for $\A_1$. Next
follows a similar forward step with respect to $\A_2$, followed by a
backward step for $\A_2$. We then proceed to the next iteration,
unless convergence is flagged.  Note that the backward steps are taken
only if they do not deteriorate the objective function beyond a
specified threshold. The procedure is repeated until a
termination condition is satisfied.

It is important to note that the method outlined above is a heuristic
extension of \cogent, and the convergence properties we have proved do
not directly translate to this setting. Nonetheless, we show below
that the method yields good empirical results.

In our first example, we consider the standard recovery of sparse +
low rank matrices. We consider a matrix of size $50 \times 50$, which
is a sum of a random rank $4$ matrix (generated by truncating the SVD
of a random Gaussian matrix) and a sparse matrix with $100$ randomly
chosen indices with standard normally distributed entries. The sets
$\mathcal{A}_1$ and $\mathcal{A}_2$ are defined in the usual way for
these types of matrices. Fig.~\ref{slr} shows that \cogent recovers
the components accurately. Note that the goal here is not to show that
\cogent outperforms other methods for Robust PCA, but rather to
demonstrate that even this problem can be solved with a variant of
\cogent.
Along similar lines, we note that \cogent can be used to deconvolve
spikes and sinusoids, or group sparse and sparse signals \cite{dirty}.

We consider now a novel application: {\em graph deconvolution}.  To
state this problem formally, consider two simple, undirected weighted
graphs $\mathcal{G}_1=(V, W_1)$ and $\mathcal{G}_2=(V, W_2)$ where $V$
represents a (common) vertex set and $W_1, W_2$ are the weighted
adjacency matrices, with superposition $W=W_1 + W_2$. Problems of this
form are of interest in \emph{covariance estimation}: $W_1$ and $W_2$
may correspond to covariance matrices of random vectors $X_1$ and
$X_2$, and from samples of $X=X_1+X_2$, one may wish to recover the
covariances $W_1$ and $W_2$.

\begin{figure}[!h]
\centering
  \includegraphics[width=.2\textwidth]{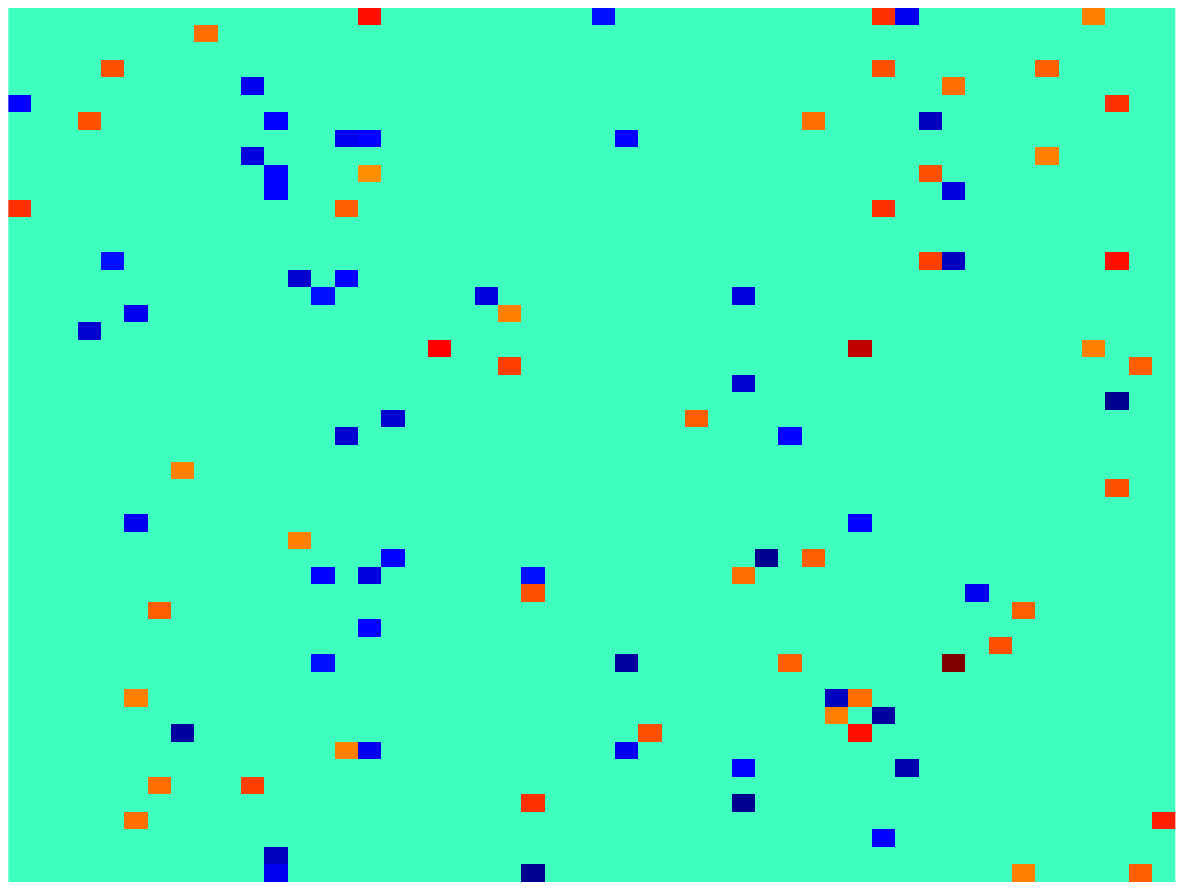}
\includegraphics[width=.2\textwidth]{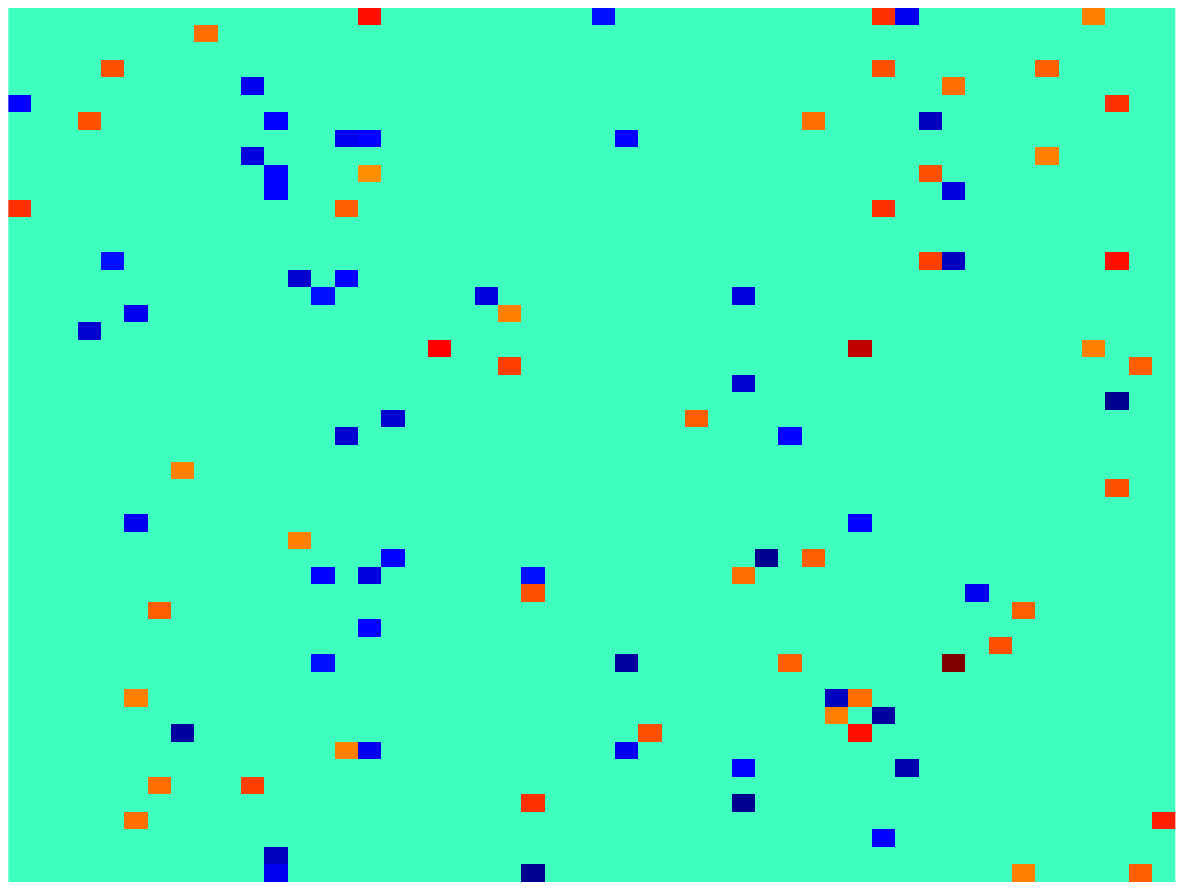} 
\vspace{-2mm}
 \includegraphics[width=.2\textwidth]{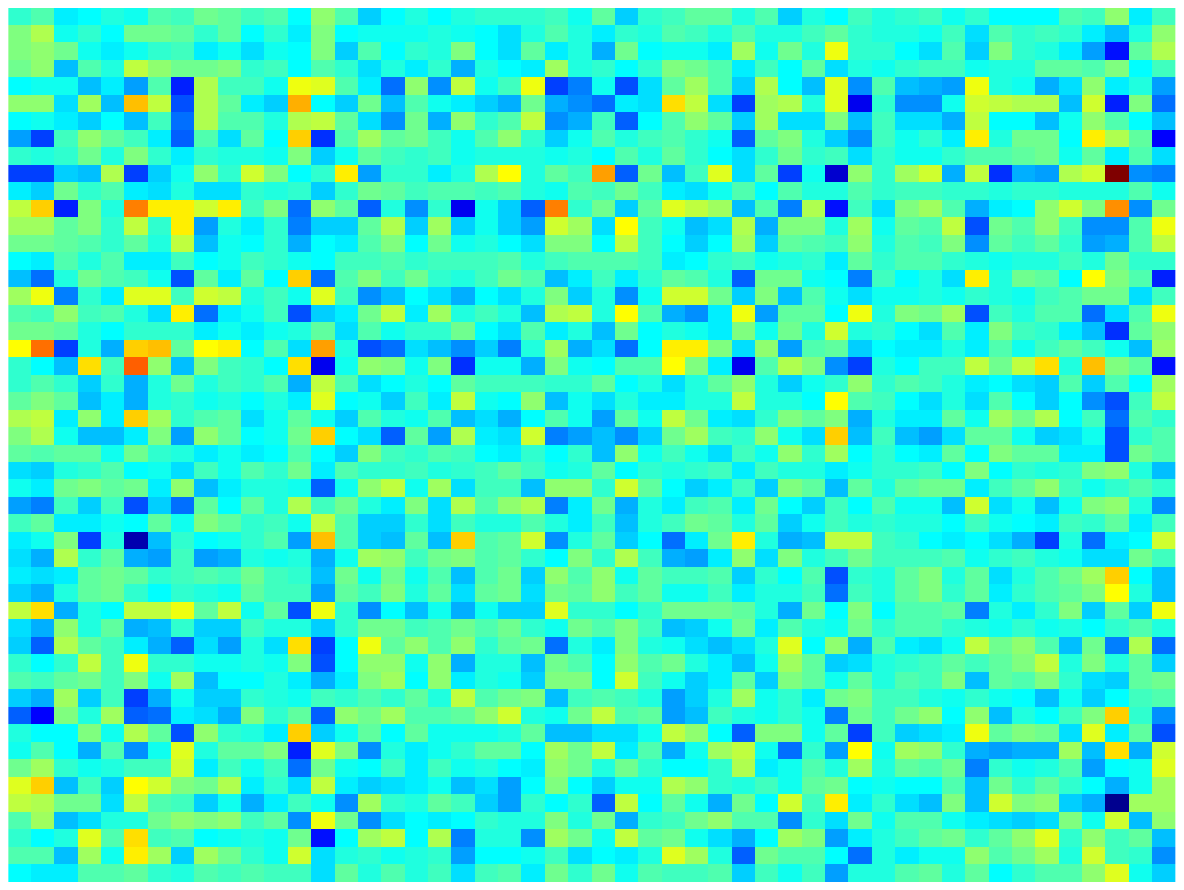}
\includegraphics[width=.2\textwidth]{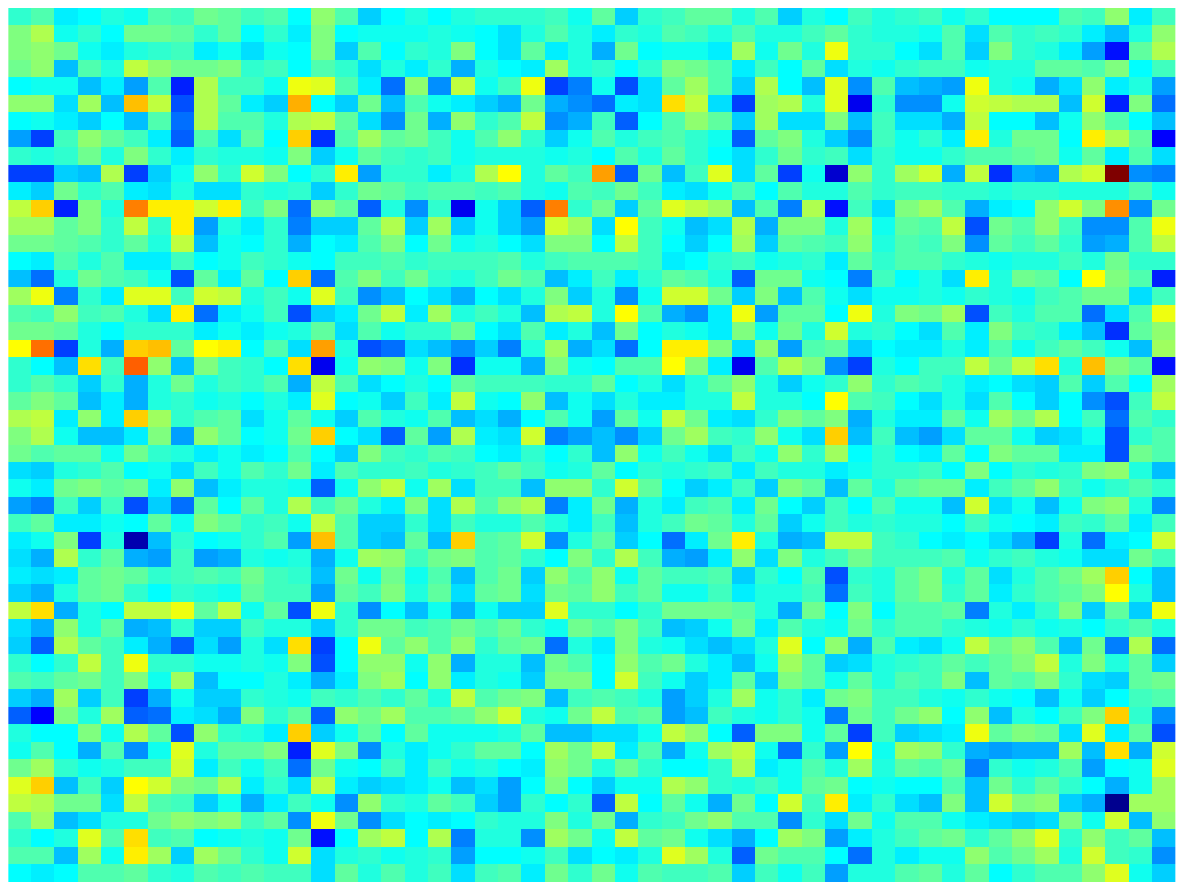} 
\caption{Recovery of a sparse + low rank matrix. The left column shows
  true components, and the right column shows recovered
  components. The top row shows the sparse part and the bottom row
  shows the low-rank part. Error in each recovered component is at
  most $10^{-7}$.}
\label{slr}
\end{figure}

%
We consider a graph of $|V|=50$ nodes in which
$\mathcal{G}_1$ and $\mathcal{G}_2$ are each restricted to a specific
family of graphs $\mathfrak{T}_1$ and $\mathfrak{T}_2$, respectively,
with the following properties.
\begin{itemize}
\item $\mathfrak{T}_1$ is the class of all tree-structured graphs on
  50 nodes. Note that the only information we exploit here is the fact
  that $\mathcal{G}_1$ is tree structured. Neither the edges of the
  tree nor the edge weights are known.
\item $\mathfrak{T}_2$ is the class of two-dimensional $5 \times 10$
  grid graphs on 50 nodes. The nodes of the graph are known up to a
  cyclic permutation. Once again, neither the edges of the graph nor
  the corresponding weights are known. The only information available
  is that one of the $50$ cyclic permutations of the nodes yields the
  desired grid-structured graph.
\end{itemize}

For set $\mathfrak{T}_1$, we define  $\A_1$ to be the
set of all matrices with Frobenius norm $1$, whose nonzero structure
is the adjacency matrix of a tree.
\footnote{We learned of the construction of tree-structured norms from
  James Saunderson, and express our gratitude for this insight.} 
For
the set $\mathfrak{T}_2$ we define  $\mathcal{A}_2$ as
follows. Let $\mathcal{P} \subseteq \R^{n \times n}$ denote the set of
all permutation matrices corresponding to the cyclic permutations
(that is, permutations in the cyclic group of order $n$). Let
$\mathcal{G}(p,q)$ (with $pq=n$) denote the set of all weighted
adjacency matrices (of unit Frobenius norm) of $p \times q$ grid
graphs with a fixed canonical labeling of the nodes. The atomic set
$\mathcal{A}_2$ is the set of weighted adjacency matrices for cyclic
permutations of all these adjacency matrices.

Given these definitions, and assuming that we observe the full
matrices, we state this deconvolution problem as:
\begin{align*}
\underset{X_1, X_2}{\text{minimize}} & \qquad  \frac12\|W-X_1-X_2\|^2 \\
\text{subject to} & \qquad \|X_1 \|_{\A_1} \leq \tau_1 ~\mbox{and} ~\|X_2 \|_{\A_2} \leq \tau_2.
\end{align*}
We need to compute the dual atomic norms to implement the forward
steps in \cogent. The variational descriptions of the dual atomic
norms are given by:
\begin{equation*}
\| Y \|_{\A_{i}}^{*} = \underset{Z \in \A_i}{\max} \left[ \text{trace}\left( ZY \right) \right]
\end{equation*}
For $\A_1$, the dual norm essentially amounts to computation of a
maximum weight spanning tree, while for $A_2$, the dual norm can be
computed in a straightforward way by sweeping through the $n$ possible
permutations of the grid graph to solve:
$$
\| Y \|_{\A_{2}}^{*} = \underset{P \in \mathcal{P}, \| \mathcal{G}(p,q) \|_F \leq 1}{\max} \; \text{trace}\left( P^{'}\mathcal{G}(p,q)PY \right).
$$ 

Our problem instances are generated as follows. We created a random
tree by generating a random (symmetric) matrix with entries
distributed as $\mathcal{U}[0,1]$, and extracting its maximum weight
spanning tree. The grid component was also chosen with similarly
chosen random weights. The resulting graphs were then superposed and
then randomly permuted. We implemented the deconvolution variant of
\cogent with backward steps as described in
Algorithm~\ref{alg:bs_multiple}. Results are shown in
Fig.~\ref{fig:graph_deconv2}.  \cogent achieves exact recovery; that
is, the edges as well as the edge weights of the constituent graphs
are correctly recovered.

\begin{figure}
\centering \subfloat[True signal is a superposition of a weighted tree
  and a weighted grid graph.]{ \centering
  \includegraphics[width=.12\textwidth]{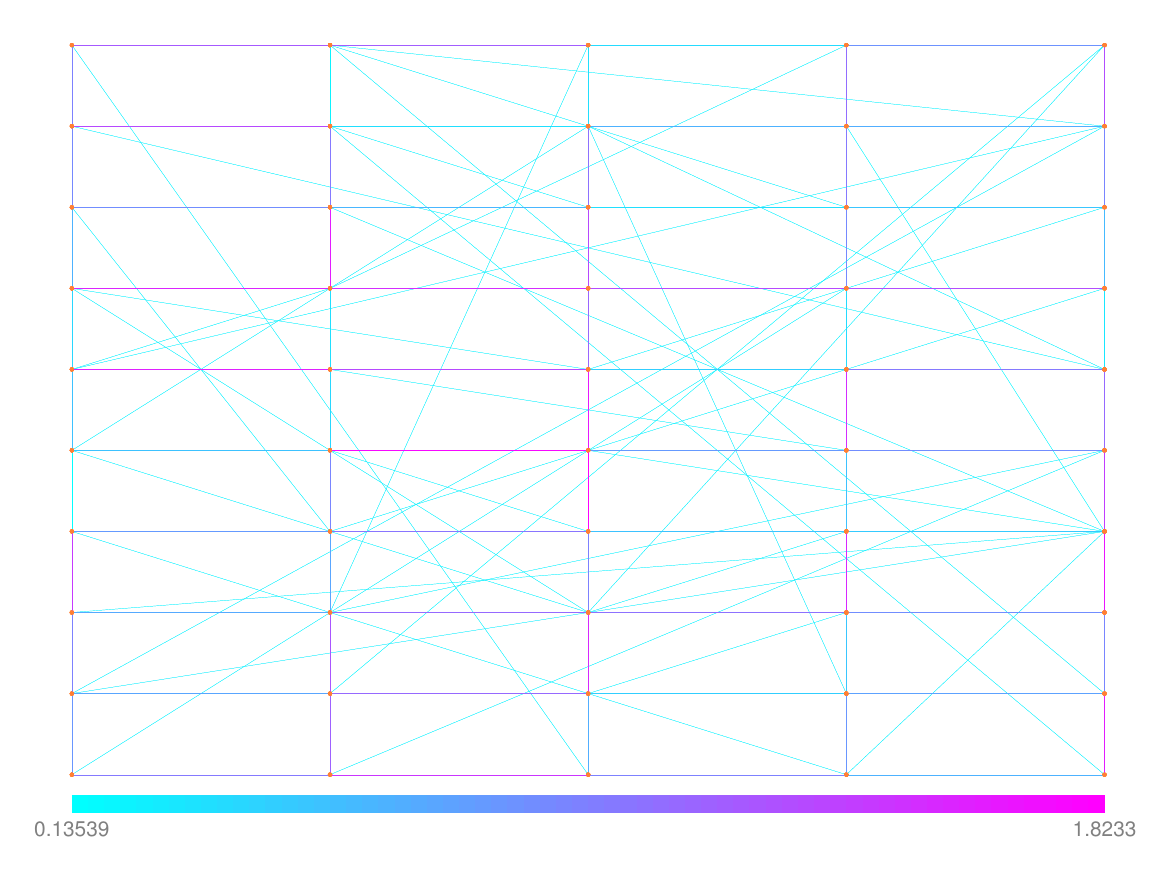}
\label{fig:graph_deconv1}
}\quad \subfloat[Tree component, recovered by \cogent.]{ \centering
  \includegraphics[width=.13\textwidth]{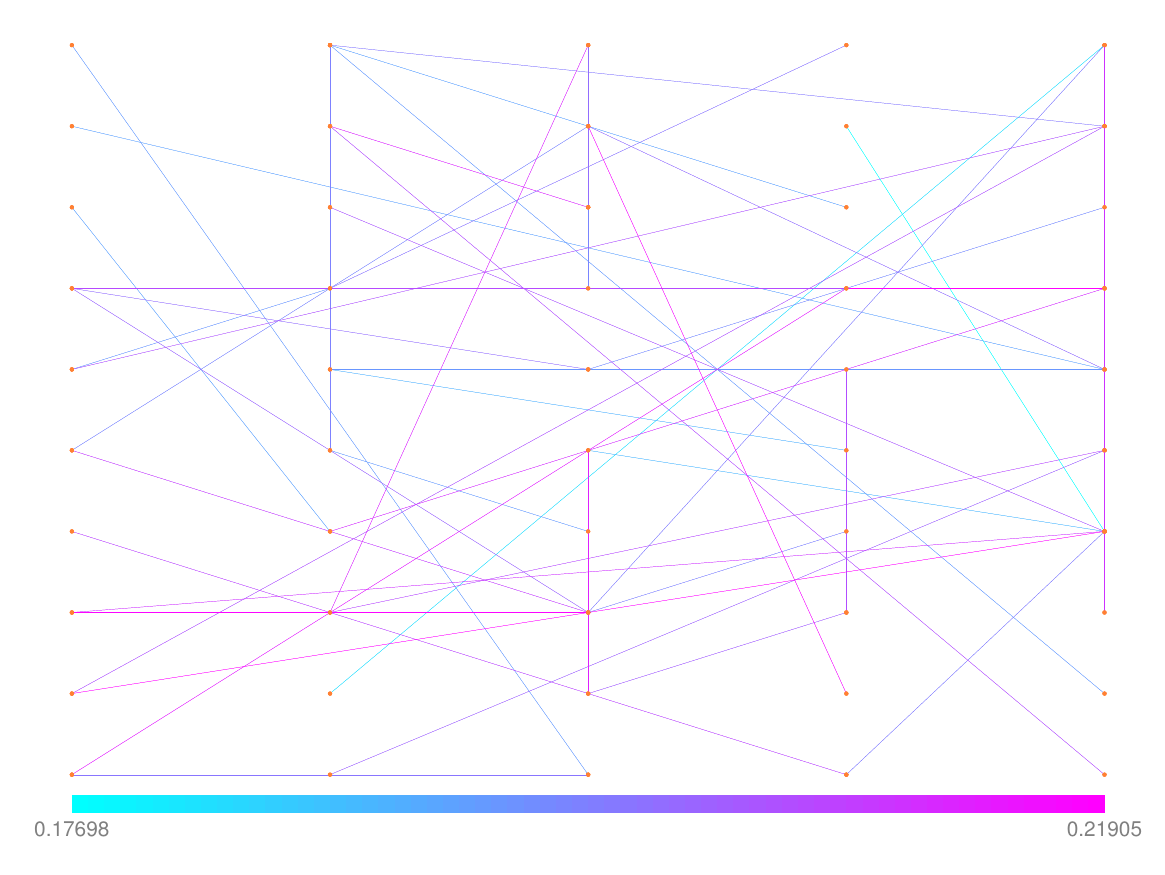}
} \quad \subfloat[Grid graph component, recovered by \cogent.]{ \centering
  \includegraphics[width=.13\textwidth]{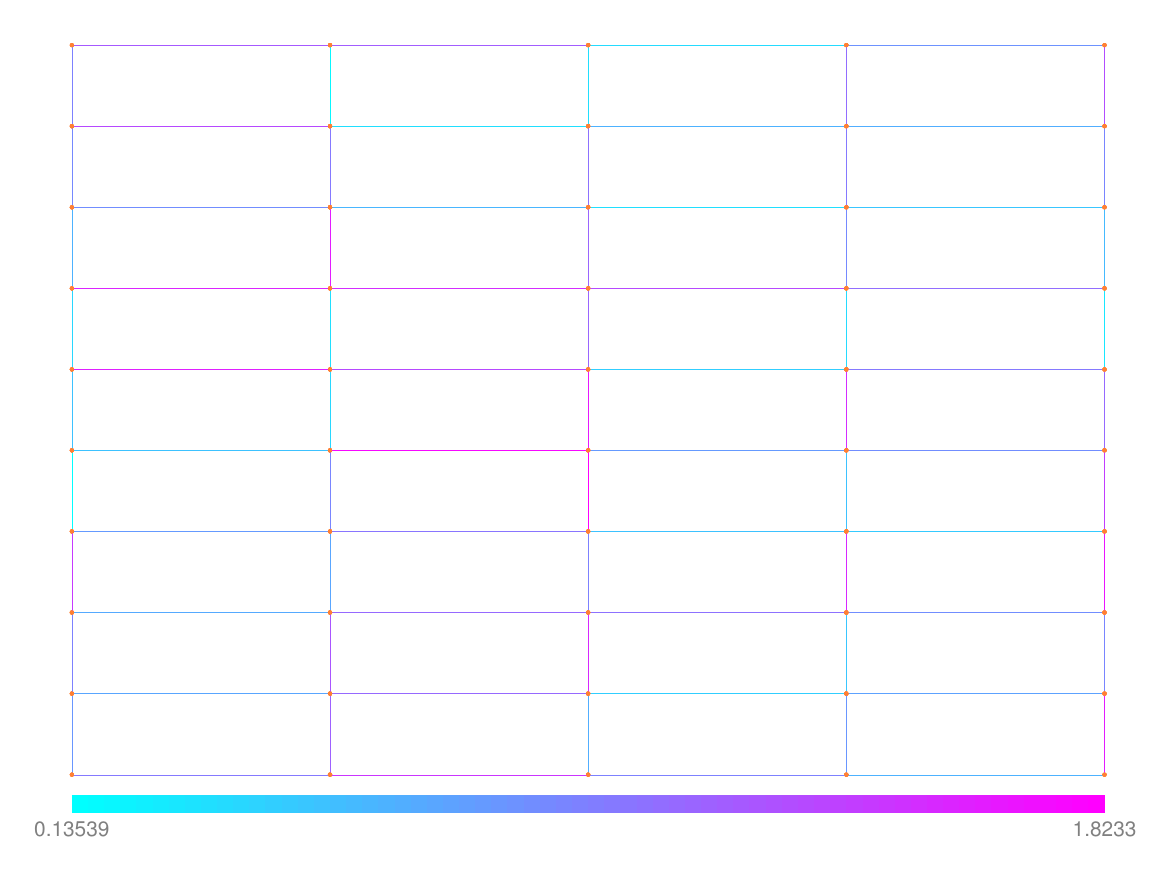}
}
\caption{Recovering constituent graph components from a superposition
  of weighted graphs. Edge weights are color-coded, with darker colors
  representing higher weights. \cogent correctly deconvolves the graph
  into its constituent components. (Best seen in color)}
\label{fig:graph_deconv2}
\end{figure}



 \section{Conclusion} \label{sec:conc}

We introduced \cogent, a greedy scheme for recovering signals that are
representable as a linear combination of a few fundamental elements
from some basis. We showed that our method is efficient and broadly
applicable, and enjoys the same theoretical convergence properties as
conditional gradient. We have described results obtained on a variety
of interesting problems, including compressed sensing, matrix
completion, moment problems and deconvolution.


\appendix 

\section{Appendix: Proofs of Convergence}

Theorem~\ref{convrateapprox} is (except for a minor difference in the
upper bounds on $\eta$) a true generalization of
Theorem~\ref{convratenoise}, in that we recover the statement of
Theorem~\ref{convratenoise} by setting $\omega=0$ in
Theorem~\ref{convrateapprox}. Likewise, the {\em proof} of
Theorem~\ref{convratenoise} can be obtained by setting $\omega=0$ in
Theorem~\ref{convrateapprox}, so we prove only the latter result here.

\subsection{Proof of Theorem~\ref{convrateapprox} }
\label{app:convrateapprox}

Denote $f_t :=f(\vx_t)$, $\tilde{f}_{t} := f(\tilde\vx_t)$, and
$f^{FW}_{t} := f(\vx_{t-1} + \gamma_{t} (\tau \va_t - \vx_{t-1}))$. We
have from the algorithm description that
\[
f_{t+1} \leq \eta f_t + (1-\eta) f^{FW}_{t+1}.
\]
For $\gamma \in [0,1]$, we   define
\[
\vx_t(\gamma) := (1-\gamma)\vx_t + \gamma \tau \va_{t+1}.
\]
Because Step~\ref{ls} of Algorithm~\ref{alg:cogent} chooses the value
of $\gamma$ optimally, we have $f^{FW}_{t+1} = f(\vx_t(\gamma_{t+1}))
\le f(\vx_t(\gamma))$, for all $\gamma \in [0,1]$, and so
\begin{align}
\notag
&f_{t+1} \\
\notag
 &\leq \eta f_t + (1-\eta) f^{FW}_{t+1} \\
\notag
& \le \eta f_t + (1-\eta) f(\vx_t(\gamma))  \\
\notag
&\le \eta f_t + \\
\notag
&(1-\eta) \left[ f_t + \nabla f(\vx_t)^T (\vx_t(\gamma) - \vx_t)\right] + \\
\notag
& (1-\eta) \left[\frac{L}{2} \| \vx_t(\gamma) - \vx_t \|^2 \right] \quad \mbox{(by definition of $L$)}\\
\notag
&= f_t + \\
\notag
&(1-\eta) \left[ \nabla f(\vx_t)^T \left( (1-\gamma)\vx_t + \gamma \tau \va_{t+1} - \vx_t \right) \right] + \\
\notag
& (1-\eta) \left[ \frac{L}{2}\| (1-\gamma)\vx_t + \gamma \tau \va_{t+1} - \vx_t \|^2  \right] \\
\notag
&= f_t + (1-\eta) \left[ \gamma \nabla f(\vx_t)^T \left( \tau \va_{t+1} - \vx_t \right) \right] + \\
\notag
& (1-\eta)\left[ \frac{L \gamma^2}{2} \| \tau \va_{t+1} - \vx_t \|^2 \right] \\
\notag
&\le f_t + (1-\eta) \left[ \gamma (1-\omega) \nabla f(\vx_t)^T (\vx^\star - \vx_t) \right] + \\
\notag
& (1-\eta) \left[2 \gamma^2 L R^2 \tau^2 \right]  \quad \mbox{(see below)}\\
\label{tosub.1}
&\le f_t + (1 - \eta) \left[ \gamma (1-\omega) (f_\star - f_t) + 2 \gamma^2 L R^2 \tau^2 \right].
\end{align}
The last inequality follows from convexity of the objective
function. The second-last inequality uses two results. First, note that
the solution $\vx^*$ can be expressed as follows:
\[
\vx^* = \sum_{\va \in \mathcal{A}} c_{\va}^* \va, \quad \mbox{for
  $c^* \ge 0$ with $\sum_{\va \in \mathcal{A}} c_{\va}^* \le
  \tau$}.
\] 
We therefore have
\begin{align*}
& \la \nabla f(\vx_t), \vx^*-\vx_t \ra  \\
& = \left\la \nabla f(\vx_t), \left( \sum_{\va \in \mathcal{A}} c_{\va}^* \va \right) -\vx_t \right\ra  \\
& \ge  \left(\sum_{\va \in \mathcal{A}} c_{\va}^* \right) \min_{\va \in \mathcal{A}} \la \nabla f(\vx_t), \va \ra -
\la \nabla f(\vx_t), \vx_t \ra \\
& \ge \min_{\va \in \mathcal{A}}  \la \nabla f(\vx_t), \tau \va - \vx_t \ra \\
& \ge \frac{1}{1-\omega} \la \nabla f(\vx_t), \tau \va_{t+1} - \vx_t \ra,
\end{align*}
by the definition of $\va_{t+1}$ in \eqref{approxgreedy} and noting
that $\min_{\va \in \mathcal{A}} \la \nabla f(\vx_t) \va \ra \leq
0$. Second, we use the definition of $R$ together with $\| \vx_t
\|_{\A} \le \tau$ and $\va_{t+1} \in \mathcal{A}$ to deduce
\[
\| \tau \va_{t+1} - \vx_t \| \le \tau \left( \| \va_{t+1} \| + \|
\vx_t/\tau\| \right) \le 2\tau R,
\]
which we can use to bound the squared-norm term.  By subtracting $f^*$
from both sides of \eqref{tosub.1}, and defining
\begin{equation} \label{eq:defdel.1}
\delta_t := f(\vx_t) - f^*,
\end{equation}
 we obtain  that 
\begin{equation} \label{eq:delrec.1}
\delta_{t+1} \le \left[ 1 - \gamma (1 - \eta) (1-\omega)  \right] \delta_t + 2 (1-\eta) LR^2 \gamma^2 \tau^2,
\end{equation} 
for all $\gamma \in [0,1]$.  This inequality implies immediately that
$\{ \delta_t \}_{t=0,1,2,\dotsc}$ is a decreasing sequence, since
$\gamma=0$ is always a valid choice in \eqref{eq:delrec.1}.

Note that $\delta_0 = f_0 - f_{\star} = D$. For the first iteration $t
= 0$, set $\gamma = 1$ in \eqref{eq:delrec.1} to obtain a further bound
on $\delta_1$:
\[
\delta_1 \leq [\eta + \omega(1- \eta)] D + 2 (1-\eta) LR^2 \tau^2 = \tilde{C}_1.
\]
For subsequent iterations $t \ge 1$, we consider the following choice
of $\gamma$:
\[
\tilde{\gamma}_t := \frac{\delta_t}{2 \tilde{C}_1}.
\]
By monotonicity of $\{ \delta_t \}$ and the bound above on
$\delta_1$, we have $\tilde{\gamma}_t \le 1/2$ for all $t \ge 1$. By
substituting the choice $\gamma = \tilde{\gamma}_t$ into
\eqref{eq:delrec.1}, we obtain
\begin{align}
\notag
\delta_{t+1} &\leq \delta_t - \delta_t^2 \frac{(1-\eta)(1-\omega) \tilde{C}_1 - (1-\eta) LR^2\tau^2}{2 \tilde{C}_1^2} \\
 \label{eq:delrec2.1}
 &= \delta_t - \frac{\delta_t^2}{\tilde{C}}.
\end{align} 
The denominator of $\tilde{C}$ is positive because $\eta
\in (0,1/3]$ and $\omega \in (0,1/4]$ together imply that
\[
(1-\omega) \tilde{C}_1 - LR^2 \tau^2  > 2(1-\omega)(1-\eta) LR^2 \tau^2 - LR^2 \tau^2 \ge 0.
\]
Note too that
\[
\tilde{C} = \frac{2 \tilde{C}_1^2}{(1-\eta) ((1-\omega)\tilde{C}_1 - LR^2 \tau^2)}  > 2 \tilde{C}_1,
\]
so that $\delta_1 \le \tilde{C}/2$.  An argument from
\cite[Lemma~2.1]{beck_cgm} yields the result.  Since $\delta_1 \le
\tilde{C}/2$, the bound \eqref{eq:convrate.approx} holds for $t=1$.  Since $\{
\delta_t \}$ is a decreasing sequence, we have $\delta_t \le
\tilde{C}/2$ for all $t \ge 1$. For the inductive step, assume that
\eqref{eq:convrate.approx} holds for some $t \ge 1$. Since the right-hand
side of \eqref{eq:delrec2.1} is an increasing function of $\delta_t$ for
all $\delta_t \in (0,\tilde{C}/2)$, this quantity can be upper-bounded
by substituting the upper bound $\tilde{C}/(t+1)$ for $\delta_t$, to
obtain
\begin{align*} 
\delta_{t+1} & \le \delta_t - \frac{\delta_t^2}{\tilde{C}} \le
\frac{\tilde{C}}{(t+1)} - \frac{\tilde{C}}{(t+1)^2} \\
& =
\frac{\tilde{C}t}{(t+1)^2} = \frac{\tilde{C}t(t+2)}{(t+1)^2 (t+2)} \le 
\frac{\tilde{C}}{t+2},
\end{align*}
establishing the inductive step and completing the proof.

\small
\bibliographystyle{IEEEtran}
\bibliography{COGENT_IEEE_tsp_Revision2_v1}

%
%
%

\end{document}